\documentclass[a4paper,11pt]{article}
\pdfoutput=1 % if your are submitting a pdflatex (i.e. if you have
\usepackage{amsmath}
\usepackage{jheppub} % for details on the use of the package, please
\usepackage[T1]{fontenc} % if needed
\usepackage{subfig}
\usepackage[ruled]{algorithm2e}

\newcommand{\nn}{\nonumber \\}

\allowdisplaybreaks[4]

\title{\boldmath Evaluating master integrals in non-factorizable corrections to $t$-channel single-top production at NNLO QCD}
\preprint{USTC-ICTS/PCFT-23-09, TTP23-011, P3H-23-016}

\author[a]{Zihao Wu,}
\author[b]{Ming-Ming Long}

\affiliation[a]{Department of Physics, University of Science and Technology of China, Hefei 230026, China}
\affiliation[b]{Institute for Theoretical Particle Physics, KIT, Karlsruhe, Germany}

\emailAdd{wuzihao@mail.ustc.edu.cn}
\emailAdd{ming-ming.long@kit.edu}

\abstract{We studied the two-loop non-factorizable Feynman diagrams for the $t$-channel single-top production process in quantum chromodynamics. We present a systematic computation of master integrals of the two-loop Feynman diagrams with one internal massive propagator in which a complete uniform transcendental basis can be built. The master integrals are derived by means of canonical differential equations and uniform transcendental integrals. The results are expressed in the form of Goncharov polylogarithm functions, whose variables are the scalar products of external momenta, as well as the masses of the top quark and the $W$ boson. We also gave a discussion on the diagrams with potential elliptic sectors.}

\begin{document} 
\maketitle
\flushbottom

\section{Introduction}
\label{sec:intro}
The top quark, the known heaviest fundamental particle, has been one of the most essential objects studied on contemporary colliders like the Tevatron and LHC since its discovery \cite{CDF:1995wbb, D0:1995jca}. Due to its substantial mass, the top quark plays a crucial part in understanding the electroweak symmetry breaking. The top quark is the only one that decays before forming a colorless bound state, or hadronization, due to its short lifetime. Given the enormous quantity of top quarks produced at the LHC, this special characteristic enables direct measurement of the top quarks' properties.

The LHC can be regarded as a top factory on which top quarks are produced in several ways\footnote{See Ref. \cite{Schwienhorst:2022yqu} for a recent review on the top physics.}. The dominant contributions come from top pair production via strong interactions and the gluon fusion channel has the largest rate. Another important way to produce top quarks is the production through electroweak interactions and a single top quark is found in the final state. Single-top production thus provides a powerful probe of the charged-current weak interactions of the top quark at hadron colliders. 

\begin{figure}[htbp]
\centering
\includegraphics[width=0.65\textwidth]{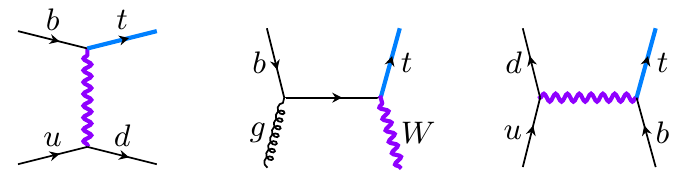}
\caption{Sample Feynman diagrams for $t$-, $tW$- and $s$- channel single-top production at tree level. }
\label{fig: tree diagrams}
\end{figure}

Single-top production may proceed via the $t$-channel, the $s$-channel, or the associated production of a top quark with a $W$ boson ($tW$ production). A space-like $W$ boson connects two quark currents in the $t$-channel single-top production. The $t$-channel dominates the single-top production and its cross section is larger than the sum of the other two production mechanisms at LHC. In the $s$-channel, the quark lines are connected by a time-like $W$ boson. It has the minimal cross section among three production channels. An on-shell $W$ boson is produced in $tW$ associated production. The sample Feynman diagrams for the three channels at the tree level are shown\footnote{In the Feynman diagrams in this paper, the blue, thick lines are for top quarks with mass $m_t$. The violet, thick, waved lines are for $W$ bosons with mass $m_W$. The black, thin, coiled lines are for gluons, which are massless. The black, thin, straight lines are for other massless quarks. The diagrams in this paper are generated using TikZ-Feynman \cite{Ellis:2016jkw}.} in Fig. \ref{fig: tree diagrams}.

The presence of charged-current weak interactions and the fact that top quarks decay almost exclusively to an on-shell $W$ boson and a $b$ quark allow for direct probe of the $tWb$ couplings in the single-top production and also for constraining the anomalous couplings in the $tWb$ vertex \cite{ATLAS:2015ryj, ATLAS:2017rcx, ATLAS:2017ygi}. Measurements of the single-top quark production cross section thus provide unbiased determinations of the essential observables like the top quark width \cite{CMS:2014mxl}, mass \cite{CMS:2017mpr}, and magnitude of the Cabibbo–Kobayashi–Maskawa (CKM) matrix element \cite{ATLAS:2019hhu, CMS:2020vac}. By studying the distributions of top-quark decay products the information about the polarization of top quark can be understood from them \cite{CMS:2015cyp, ATLAS:2022vym}. The single-top production also plays a vital role in providing interesting probes of parton distribution functions (PDFs) and searching for new physics signals at the LHC \cite{ATLAS:2017rso, CMS:2019jjp}.

The theoretical studies on single-top production over the decades have built a solid foundation for understanding the electroweak properties of the top quark. The next-to-leading order (NLO) QCD corrections to all three channels of single-top production have been known for decades \cite{Harris:2002md,Zhu:2002uj}. The next-next-to-leading order (NNLO) QCD corrections to $t$- and $s$- channel single-top production, in the structure function approximation, are computed in Refs. \cite{Brucherseifer:2014ama,Berger:2016oht,Berger:2017zof,Campbell:2020fhf} and Ref. \cite{Liu:2018gxa}, respectively. Previous studies neglect the talk between heavy and light quark lines, i.e. the non-factorizable contributions, for their small corrections. However, it is argued in Refs. \cite{Bronnum-Hansen:2021pqc,Bronnum-Hansen:2022tmr} recently that the non-factorizable contributions to $t$-channel single-top production will be enhanced by a factor $\pi^2$ and thus can not be neglected. The complicated multi-scale two-loop Feynman integrals in Refs. \cite{Bronnum-Hansen:2021pqc,Bronnum-Hansen:2022tmr} are computed numerically by applying the auxiliary mass flow method \cite{Liu:2017jxz,Liu:2020kpc,Liu:2021wks}. As for the $tW$ mode, the complete corrections at NNLO QCD are still unavailable. Only partial results have been obtained \cite{Chen:2021gjv,Long:2021vse,Chen:2022ntw,Chen:2022yni,Wang:2022enl} and the complete two-loop QCD amplitudes for $tW$ production were studied very recently \cite{Chen:2022pdw}.

We aim to perform an independent calculation of the non-factorizable contributions to $t$-channel single-top production. A fast and stable evaluation of amplitudes during phase space integration is demanded in the phenomenological study. This evaluation will be much easier if the building blocks of the amplitudes, the Feynman integrals, can be expressed by some special functions, which are well understood and easily and precisely evaluated. Being different from Refs. \cite{Bronnum-Hansen:2021pqc,Bronnum-Hansen:2022tmr}, we try to calculate the relevant two-loop Feynman integrals analytically which is a challenging task due to the many scales. 

The two-loop amplitude contribution consists of 18 non-factorizable diagrams, including planar and non-planar diagrams. The various Feynman integrals on each diagram can be reduced as a linear combination of a relatively smaller number of selected integrals. Thus, for a certain diagram, only these independent integrals, called \textit{master integrals}, need to be computed analytically. The technique of the master integral computation varies. Among them, one of the most useful methods is based on the differential equations of uniform transcendental (UT) integral basis \cite{Henn:2013pwa,Henn:2014qga}.  This method derives the selected UT master integrals by solving the canonical differential equations on the kinematic variables, via iterative integrations. Using this method, the integrals can be expressed in form of a Laurent expansion of the spacetime parameter $\epsilon$ with coefficients being transcendental functions, such as Goncharov polylogarithm (GPL) functions. In this paper, we are trying to use the UT basis method to evaluate the master integrals of the above diagrams. We gave the results of all the diagrams with one internal massive propagator, including a planar diagram and two non-planar diagrams. We expressed their results in GPL functions. We also gave a general discussion on the other diagrams. 

We remark that, during the preparation of this paper, researchers working on the same problem have made their progress \cite{Syrrakos:2023syr}. In this work, the method named Simplified Differential Equations is applied, and the master integrals for a planar diagram and a non-planar diagram are evaluated. These two diagrams are also included in our computation.

Usually, one needs to validate his or her analytic results of Feynman integrals after the analytic computation. There are different kinds of validation methods. One convenient way is to make use of numerical Feynman integral evaluation tools, including AMFlow \cite{Liu:2022mfb,Liu:2022chg,Liu:2021wks,Liu:2020kpc,Liu:2018dmc,Liu:2017jxz}, {\sc Fiesta} \cite{Smirnov:2021rhf,Smirnov:2015mct,Smirnov:2013eza,Smirnov:2009pb,Smirnov:2008py}, SecDec (and pySecDec) \cite{Borowka:2017idc,Borowka:2015mxa,Jahn:2018zsh}. The analytic results should agree with the numerical ones if derived correctly. 

This paper is organized as follows. In section \ref{sec:review}, we provide a review of the methodology of the integration-by-parts reduction, the canonical differential equations, and the uniform transcendental integrals. In section \ref{sec:diagram}, we provide an analysis of the NNLO Feynman diagrams for non-factorizable corrections to $t$-channel single-top production. We also provide our computation and corresponding results of the diagrams with one internal mass in this section. These diagrams are free of elliptic sectors. In section \ref{sec:summary}, we summarize this paper.

\section{Review of methods}\label{sec:review}
In this section, we are reviewing the evaluation method we are using in this paper, including the integration-by-parts reduction, uniform transcendental integrals, and canonical differential equations.

\subsection{The integration-by-parts reduction}
The Feynman integrals of a diagram can be reduced to linear combinations of scalar integrals after the Passarino-Veltman reduction \cite{Passarino:1978jh}. The scalar integrals, which form a linear space called a Feynman integral family, can be expressed in the standard form as
\begin{equation}
    G_{\alpha_1,\dots,\alpha_n}:=\int \frac{{\rm d}^D l_1}{i \pi^{D/2}}\cdots\frac{{\rm d}^D l_L}{i \pi^{D/2}}\frac{1}{D_1^{\alpha_1}\cdots D_n^{\alpha_n}},
\end{equation}
where $L$ is the loop number and $D_i$'s are the propagators of the given integral family. In this Feynman integral family, there exist linear constraints called the integration-by-parts (IBP) relations \cite{Tkachov:1981wb,Chetyrkin:1981qh}. The IBP relations are given by the derivative of certain integrals, which vanish in the sense of dimensional regularization, as
\begin{equation}\label{eq:IBP}
    0=\int \frac{{\rm d}^D l_1}{i \pi^{D/2}}\cdots\frac{{\rm d}^D l_L}{i \pi^{D/2}}\frac{\partial}{\partial l^\mu}\frac{v^\mu}{D_1^{\alpha_1}\cdots D_n^{\alpha_n}},
\end{equation}
where $v^\mu$ is a chosen vector being a combination of external or loop momenta. The right-hand side of \eqref{eq:IBP} expands to a linear combination of Feynman integrals. This means that the integrals in a family are not linearly independent and constrained by IBP relations. Given a target integral in the family, using the IBP relations, it can be reduced as a linear combination to a chosen basis made by finite (\cite{Smirnov:2010hn}) number of integrals, as
\begin{equation}\label{eq:IBP reduction}
    I_{\text {target}}=\sum_{i=1}^N c_i I_i,
\end{equation}
where $I_i$'s are the chosen, linearly independent integrals, called master integrals, and $c_i's$ are called IBP reduction coefficients. Reducing a target integral in form as the right-hand side of \eqref{eq:IBP reduction} is called IBP reduction. IBP reduction can be performed via some released computer programs or packages, such as AIR, FIRE, FiniteFlow, {\sc Kira}, {\sc LiteRed} and {\sc Reduze} \cite{Anastasiou:2004vj,Smirnov:2008iw,Smirnov:2013dia,Smirnov:2014hma,Smirnov:2019qkx,Peraro:2019svx,Maierhofer:2017gsa,Maierhofer:2018gpa,Maierhofer:2019goc,Lee:2013mka,Studerus:2009ye,vonManteuffel:2012np}, etc. Besides, there are packages that can be used to simplify reduction coefficients, including pfd-parallel and MultivariateApart \cite{Heller:2021qkz,Bendle:2021ueg,Boehm:2020ijp}. In our work, we mainly use FIRE6 \cite{Smirnov:2019qkx} to perform IBP reductions.

The procedure of computing target integrals is with two submissions. One is to perform IBP reduction \eqref{eq:IBP reduction} to get the IBP reduction coefficients $c_i$, as stated in the last paragraph. The other is to compute the analytic expressions of master integrals. One can use the methods like Feynman representation or Mellin-Barnes \cite{Smirnov:1999gc, Tausk:1999vh} and expansion-by-regions \cite{Beneke:1997zp} techniques to compute the master integrals analytically. Their power is limited because of the involving complicated parametric integrations. In this paper, we are using the method of differential equations and the formerly mentioned ones will serve as the auxiliary aid when determining the boundary conditions of the differential equations. In general, Feynman integrals are functions of kinematic variables, labeling as $x_i$, and spacetime dimension parameter $\epsilon$, defined as $d=4-2\epsilon$, where $d$ is the spacetime dimension. The derivative of Feynman integrals with respect to $x_i$'s are also combinations of Feynman integrals in the same family. After IBP reduction, they can be written as linear combinations of master integrals. Thus, for master integrals, we have
\begin{equation}\label{eq:DE}
    \frac{\partial}{\partial x_i}I = (A_i)I,
\end{equation}
where $I$ denotes the column vector formed by master integrals, and matrices $A_i$'s are differential equation matrices with respect to $x_i$. In general, $A_i$'s are functions of $x_i$'s and $\epsilon$, i.e. $A_i=A_i(x_j,\epsilon)$. 

\subsection{Uniform transcendental integrals and canonical differential equations}

Solving differential equations like \eqref{eq:DE} is not easy. However, for many diagrams, we are able to find a better choice of master integral basis called uniform transcendental (UT) basis \cite{Henn:2013pwa,Henn:2014qga}, a kind of basis made of UT integrals. The definition of UT integral is that the coefficients of $\epsilon$-expansion of UT integrals are with a transcendental weight that matches the power of $\epsilon$. Usually, for $L$-loop UT integrals $I_i$, we have
\begin{equation}\label{eq:epsilon expansion}
    I_i=\epsilon^{-2 L}\sum_{k=0}^\infty I_i^{(k)} \epsilon^k,
\end{equation}
where $I_i^{(k)}$ are transcendental functions with weight $k$. So far, we have not explained what transcendental weight is. In general, a rational function whose coefficients are rational numbers $f^{(0)}$ can be considered as a weight-0 transcendental function, written as $\mathcal T(f^{(0)})=0$. Weight-$k$ functions are related to functions with lower weight by 1 as
\begin{equation}
    \mathcal T (\int f^{(k)} \text{dlog}f^{(0)})=k+1,
\end{equation}
where $f^{(k)}$ and $f^{(0)}$ are weight-$k$ and weight-0 functions respectively. For example, we have $\mathcal{T} (\log (x))=1$ and $\mathcal T(\text{Li}_2(x))=2$ since 
\begin{equation}
    \text{Li}_2(x)=-\int_0^x \text{d}x^\prime \log(1-x^\prime) \text{dlog}(x^\prime).
\end{equation}
Some irrational numbers may be of weight higher than 0 since they can be considered as corresponding transcendental functions, such like $\mathcal T(\pi)=\mathcal T(\log(-1))=1$.

After choosing a UT basis, the differential equation is proportional to $\epsilon$ as
\begin{equation}\label{eq:UTDE}
    \frac{\partial}{\partial x_i}I = \epsilon A_i(x_j)I,
\end{equation}
where $A_i$ is called the matrices of canonical differential equations, whose entries are usually rational functions (or at most algebraic functions with some square roots). With the parameter $\epsilon$ factored out, we can perform integration on \eqref{eq:UTDE} to derive the analytic results of the UT integrals in that basis, in form of $\epsilon$-expansion shown in \eqref{eq:epsilon expansion}.

There is another very useful property of canonical differential equation matrix $A_i(x)$. Considering that $[\partial_{x_i},\partial_{x_j}]I=0$, together with \eqref{eq:UTDE}, we have
\begin{equation}\label{eq:integrability of UTDE}
    \frac{\partial}{\partial x_i}A_j-\frac{\partial}{\partial x_j}A_k=0,
\end{equation}
and
\begin{equation}
    [A_i,A_j]=0.
\end{equation}
Eq. \eqref{eq:integrability of UTDE} shows that there exists a matrix $\tilde{A}$ such that
\begin{equation}\label{eq:Ai and Atilde}
    A_i=\frac{\partial}{\partial x_i}\tilde{A}.
\end{equation}
The other requirement about the canonical differential equations is that they must be in a Fuchsian form and, more concretely, a $d\text{log}$ form, as follows 
\begin{equation}\label{eq:Atilde in letters}
    \Tilde{A}=\sum_i a_i \log W_i,
\end{equation}
where $a_i$'s are matrices of rational numbers and $W_k$'s, called \textit{symbol letters}, are rational or algebraic functions of the kinematic variables. They are closely related to the intrinsic singularities and branch cuts of the integrals in the Feynman diagram.

\subsection{UT basis determination}
Before deriving the canonical differential equations, we need to find a UT basis first. Nowadays, there are many methods to find UT basis. Theoretical methods include leading singularity analysis \cite{Henn:2013pwa,Henn:2014qga,Dlapa:2021qsl,He:2022ctv}, $d\text{log}$ integrals \cite{Wasser:2018qvj,Chicherin:2018old,Henn:2020lye}, intersection theory \cite{Frellesvig:2020qot,Chen:2022lzr,Chen:2022fyw,Chen:2020uyk}, Magnus series \cite{Argeri:2014qva}, etc. Besides, there are some algorithms and packages for UT integral determination, they are: {\sc Canonica} \cite{Meyer:2016slj,Meyer:2017joq}, \texttt{DlogBasis} \cite{Henn:2020lye}, {\sc epsilon} \cite{Prausa:2017ltv}, {\sc Fuchsia} \cite{Gituliar:2017vzm},  {\sc initial} \cite{Dlapa:2020cwj}, {\sc libra} \cite{Lee:2014ioa,Lee:2017oca,Lee:2020zfb}, etc. These methods are with different kinds of advantages.
Usually, one often needs to try different methods for a given UT determination problem.

Here we explain the main method we use in our work on the problem involved in this paper.  Given an integral family, it is not guaranteed that we can use the methods introduced above to generate a linearly complete UT basis. But it is important to use those methods to find as many linear independent UT integrals as possible. Afterwards, we have fewer UT integrals to be found, manually, to form a complete linear basis. Fortunately, with the help of the concept of \textit{sector}, the manual procedure can be made easier.

 The sector of a given integral is describing whether a propagator appears as a denominator (labeled 1) or numerator (labeled 0) in the integrand, i.e. the sector of an integral $G_{\alpha_1,\cdots,\alpha_n}$ is $(a_1,\cdots,a_n)$ where
\begin{equation}
a_i=\left\{
             \begin{array}{lr}
             1,\quad\alpha_i>0\\
             0,\quad\alpha_i\leq0
             \end{array}
\right.
\end{equation}
Altering one or more 1 to 0 in the indices of a sector A gives its \textit{sub-sector} B, which means eliminating the corresponding propagator in the denominator. Equivalent speaking, we state that A is a \textit{super-sector} of B. In general, if an integral is a linear combination of integrals, which corresponds to sector A and possibly A's sub-sectors, we say that this integral is in sector A.

Suppose we already have several UT integrals $I^\prime$ derived from the existing methods, we need firstly to determine which sectors the rest UT integrals are from, and to complete the subset of UT basis in this sector manually. This can be done by finding the subset of the master integrals that can form an independent and complete linear space together with $I^\prime$. The choice of such a subset is not unique. A good choice is to let the sectors of the master integrals chosen to be from lower sectors, which means, with less propagators in the denominator. The UT integrals in these sectors are either known or easy to determine, by introducing double (or possibly triple) propagators \cite{Henn:2014qga}.

This sector-by-sector basis completing method can be used if $I^\prime$ is enough to form a nearly-complete basis. Sometimes, it is not the case. We will encounter this circumstance in section \ref{subsec:rxb1}, where we found another way to determine the UT basis using Magnus series.

The Magnus series method requires that we have already obtained a linear basis. The meaning of a linear basis is that its corresponding differential equation matrix is linear in $\epsilon$, namely, 
\begin{equation}
    \frac{\partial}{\partial x_i} J = (A_{i,0}+\epsilon A_{i,1})J.
\end{equation}
Suppose that the UT basis is related to $J$ by $J=\text{T}I$, where $T$ is free of $\epsilon$. Then, finding the desired transformation matrix $\text{T}$ reduces to solve the following matrix equation,
\begin{equation}
    \frac{\partial}{\partial x_i} \text{T} = A_{i,0}\text{T}.
\end{equation}
The solution to the above equation can be expressed by the exponential of a series as
\begin{equation}
    \text{T}=e^{ \Omega[A_{i,0}]},
\end{equation}
where
\begin{equation}\label{eq:Magnus series}
    \Omega[A_{i,0}]:= \sum_{n=1}^{\infty}\Omega_n[A_{i,0}],
\end{equation}
which is called the Magnus expansion, and 
\begin{equation}\label{eq:expansions of Magnus series}
\begin{aligned}
\Omega_{1}[A(t)] &= \int^{t} d z_{1} A(z_{1}), \\
\Omega_{2}[A(t)] &= \frac{1}{2} \int^{t} d z_{1} \int^{z_{1}} d z_{2}\left[A(z_{1}), A(z_{2})\right], \\
\Omega_{3}[A(t)] &= \frac{1}{6} \int^{t} d z_{1} \int^{z_{1}} d z_{2} \int^{z_{2}} d z_{3}
\Bigg\{
	\left[A(z_{1}),\left[A(z_{2}), A(z_{3})\right]\right] \\
	&\hspace{6em} +\left[A(z_{3}),\left[A(z_{2}), A(z_{1})\right]\right]
\Bigg\},\\
\vdots
\end{aligned}
\end{equation}

Notice that this method works well only if the Magnus series \eqref{eq:Magnus series} terminates at a certain order. If so, one can acquire a nice transformation matrix from \eqref{eq:Magnus series}. This can be done in a systematic way. The corresponding workflow can be designed as in Algorithm \ref{algorithm: Magnus series}, where $N$ denotes the number of $x_i$'s and the subscripts "diag" and "nd" denote the diagonal and non-diagonal parts of the matrices, respectively. Note that in the transformation of the diagonal part, the Magnus series terminates at order 1 since all commutators vanish in \eqref{eq:expansions of Magnus series}. Finally, the product 
\begin{equation}
    \text{T}_1 \cdots \text{T}_{2N}
\end{equation}
is the matrix which transforms the original differential equation into the canonical form.

\begin{algorithm}[htbp]
 \SetAlgoLined
 \LinesNumbered
 \SetKwInOut{Input}{Input}
 \SetKwInOut{Output}{Output}
 
 \Input{$\frac{\partial}{\partial x_i} J^{(0)} = \left(A_{i,0}^{(0)}+\epsilon A_{i,1}^{(0)}\right)J^{(0)}$}
 \tcp{\emph{Transforming diagonal part}}
\For{$n\leftarrow 1$ \KwTo $N$}{
$\text{T}_{n,{\text {diag}}} = e^{\Omega_1[A^{(n-1)}_{n,0,{\text {diag}}}]}$ \;
$J^{(n)} = \text{T}^{-1}_{n,{\text {diag}}}J^{(n-1)}$  \;
$A_i^{(n)} = \text{T}_{n,{\text {diag}}}A_i^{(n-1)}\text{T}_{n,{\text {diag}}}^{-1}-\text{T}_{n,{\text {diag}}}\frac{\partial}{\partial x_i}\text{T}_{n,{\text {diag}}}^{-1}= A^{(n)}_{i,0,{\text {nd}}} + \epsilon A^{(n)}_{i,1}$\;
}
\tcp{\emph{Transforming non-diagonal part}}
\For{$n\leftarrow N+1$ \KwTo $2N$}{
$\text{T}_{n,{\text {nd}}} = e^{\Omega[A^{(n-1)}_{n,0,{\text {nd}}}]}$ \;
$J^{(n)} = \text{T}^{-1}_{n,{\text {nd}}}J^{(n-1)}$ \;
$A_i^{(n)} = \text{T}_{n,{\text {nd}}}A_i^{(n-1)}\text{T}_{n,{\text {nd}}}^{-1}-\text{T}_{n,{\text {nd}}}\frac{\partial}{\partial x_i}\text{T}_{n,{\text {nd}}}^{-1}= \epsilon A^{(n)}_{i,1}$
}
\Output{$\frac{\partial}{\partial x_i} I = \epsilon  A^{(2N)}_{i,1}I$ and $\text{T}_1 \cdots \text{T}_{2N}$}
\caption{Transformation by Magnus Series}
\label{algorithm: Magnus series}
\end{algorithm}

\subsection{Solving the canonical differential equations}\label{sec:solving DE}
From \eqref{eq:UTDE} and \eqref{eq:Ai and Atilde}, the differential equations can be written as the $d\text{log}$ form,
\begin{equation}\label{eq:dlog form}
    d I = \epsilon (d \tilde{A}) I.
\end{equation}
Substituting the expansion of $I$
\begin{equation}
    I = \epsilon^{-2L} \sum_{k=0}^{\infty} I^{(k)}\epsilon^k
\end{equation}
back into the $d\text{log}$ form \eqref{eq:dlog form}, one can build a recursive relation among different $I^{(k)}$'s,
\begin{equation}
    d I^{(k)} = 
    \left\{
    \begin{array}{lr}
        \quad 0 & k=0\\
        (d\tilde{A}) I^{(k-1)} & k > 0
    \end{array}
    \right. .
\end{equation}
It is straightforward to see that $I^{(k)}$ can be determined order by order in $\epsilon$,
\begin{align}
I^{(0)}(x) &= I^{(0)}(x_b)\,, \\
I^{(1)}(x) &= I^{(1)}(x_b)
    +\int_{\gamma}d\tilde{A}I^{(0)}(x_b)\,, \\
I^{(2)}(x) &= I^{(2)}(x_0)
    +\int_{\gamma}d\tilde{A}I^{(1)}(x_b)
    +\int_{\gamma}d\tilde{A}d\tilde{A}I^{(0)}(x_b)\,, \\
  & \vdots\nonumber
\end{align}
where $\gamma$ is a piecewise smooth path that connects the boundary $x_b$ and a general $x$ in the parameter space. Then the solution is expressed by Chen's iterated integrals \cite{Chen:1977oja}. The integration results are independent of the choice of path $\gamma$ unless the path variation crosses a singular point with nonzero residue. A convenient choice of $\gamma$ is the composed path of consequent segments with each along one of the axes in the parameter space. Without loss of generality, let the integration be performed firstly in $x_1$, the solution is fixed up to an undetermined function of remaining $x_i$'s. Substituting the obtained solution into the differential equation of $x_2$, one can get a new differential equation of the undetermined function in the last step with respect to $x_2$. Integrating the new differential equation will update the solution with explicit dependence on $x_2$. Repeating similar procedures on following $x_i$'s, one eventually gets the complete solution up to an unknown constant vector at each order in $\epsilon$. Even though the solution is, without crossing singular point(s) with a nonzero residue, invariant under different $\gamma$, the same path should be chosen consistently in the recursive calculation.

In the special case when the integration kernel takes the general $d\log$ form
\begin{equation}
    \frac{dx_i}{x_i-r},
\end{equation}
where $r$ is an algebraic function of the remaining variables, the solution straightforwardly evaluates to the Goncharov polylogarithms (GPLs) \cite{Goncharov:2001iea, Goncharov:1998kja} which are a class of special Chen's iterated integral. A wight $n$ GPL with indices $w_i~ (i = 1, ..., n)$ and argument $t$ is defined recursively by
\begin{equation}
G(w_n, ..., w_1; t) = \int_0^t \frac{1}{\tau - w_n}\,G(w_{n-1}, ..., w_1; \tau)\,d\tau
\label{eq:GPLdef}
\end{equation}
with $G(\,; t) = 1$ and
\begin{equation}
G(\vec{0}_n; t) = \frac{\log^n t}{n!}\,.
\end{equation}

The final step is to fix the unknown constant vector at each order in $\epsilon$ with the help of boundary conditions. There are several ways to do it. The simplest case is that some of the UT integrals, usually those with less propagators, can be computed at a particular kinematic point in the exact or asymptotic form by methods like Feynman representation, Mellin-Barnes \cite{Smirnov:1999gc, Tausk:1999vh}, and expansion-by-regions \cite{Beneke:1997zp} techniques. For the other integrals, another approach is to make use of the characters of UT integrals, especially the behavior around singularities. Usually, the singularities that appear in the canonical differential equations contain two types, physical singularities or spurious singularities. For the latter ones, no divergence would actually show up when reaching those spurious singularities. Thus, the linear relations among constants of distinct UT integrals can be constructed systematically given the knowledge of whether the integrals are singular or not in some limits. The starting step is to determine which singularities are physical and which are spurious. This can be analyzed in two ways. The first is to perform the region analysis under the interested limits. If only the hard region is revealed, then it's safe to sentence the regularity of an integral in that specific limit\footnote{Because a UT integral is often composed of several integrals, it's not rigorous to judge the divergence even if each term is singular. Magic cancellation might happen. A solid conclusion will take more investigation. One can just give up such a potential contribution to boundary constant determination because other unambiguous conclusions are always sufficient (in our calculation).}. The second way is to notice the physical singularities are encoded in the logarithms that appear at weight 1 of each UT integral. One can easily fit the weight 0 constant vector from a quick numerical evaluation with the help of pySecDec \cite{Borowka:2017idc} or AMFlow \cite{Liu:2022chg} since it is constituted of rational numbers. Then the weight 1 of every UT integral can be obtained by integrating the canonical differential equation in the mentioned manner. These two methods provide a cross-check with each other in our calculation. In our work, the above two methods, direct evaluation, and spurious singularity constraints are used together to provide boundary conditions in each integral family.

\section{The Feynman diagrams and the UT integrals}
\label{sec:diagram}
We first briefly present the setup for our later calculations of the two-loop master integrals that contribute to the $t-$channel single-top production. Let's consider the following partonic scattering process,
\begin{equation}
    u(p_1) + b(p_2) \rightarrow t(-p_3) + d(-p_4),
\end{equation}
where only $t$ is massive, i.e. $p_{1,2,4}^2=0$ and $p_3^2=m_t^2$. With three independent momenta $p_{1,2,4}$, three Mandelstam variables can be constructed as
\begin{equation}
    s = (p_1+p_2)^2, \quad
    t = (p_1+p_4)^2, \quad
    u = (p_2+p_4)^2.
\end{equation}
They sum up to the square of the top's mass. In the following sections, we will choose $s,t$ as the independent ones. Together with $m_t^2$ and $m_W^2$, which comes from the internal line, the master integrals to be calculated are the functions of $s,t,m_t^2$ and $m_W^2$. Note that for single-top production, we have
\begin{equation}
    s > m_t^2 > m_W^2 >0, \quad m_t^2-s < t <0. 
\end{equation}

\begin{figure}[htbp]
\centering
\includegraphics[width=0.45\textwidth]{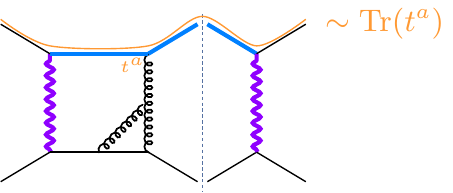}
\caption{Color conservation reduces the number of diagrams for non-factorizable corrections to the $t$-channel single-top production.}
\label{fig:color conserve}
\end{figure}

\begin{figure}[htbp]
  \centering
  \captionsetup[subfigure]{labelformat=empty}
  \subfloat[$\mathcal{T}_1$]{%
    \includegraphics[width=0.3\textwidth]{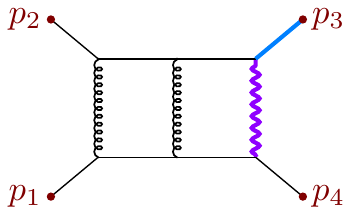}
  }
  \subfloat[$\mathcal{T}_2$]{%
    \includegraphics[width=0.3\textwidth]{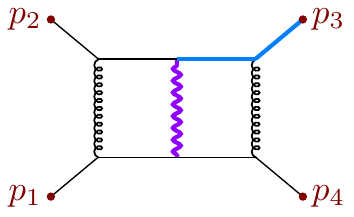}
  }
  \subfloat[$\mathcal{T}_3$]{%
    \includegraphics[width=0.3\textwidth]{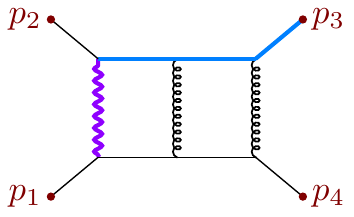}
  }
  \\
  \subfloat[$\mathcal{T}_4$]{%
    \includegraphics[width=0.3\textwidth]{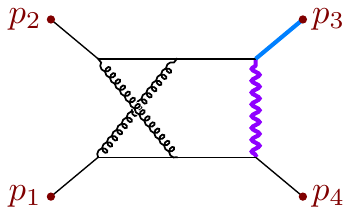}
  }
  \subfloat[$\mathcal{T}_5$]{%
    \includegraphics[width=0.3\textwidth]{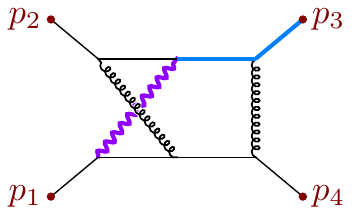}
  }
  \subfloat[$\mathcal{T}_6$]{%
    \includegraphics[width=0.3\textwidth]{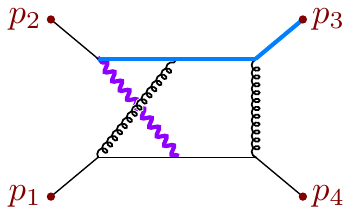}
  }
  \\
  \subfloat[$\mathcal{T}_7$]{%
    \includegraphics[width=0.3\textwidth]{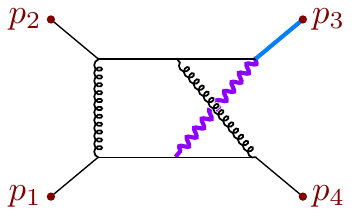}
  }
  \subfloat[$\mathcal{T}_8$]{%
    \includegraphics[width=0.3\textwidth]{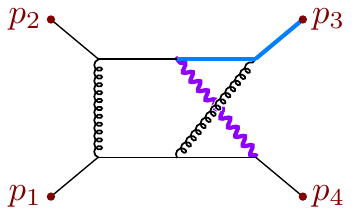}
  }
  \subfloat[$\mathcal{T}_9$]{%
    \includegraphics[width=0.3\textwidth]{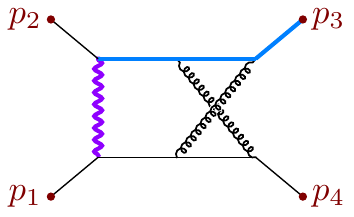}
  }
\caption{The non-factorizable Feynman diagrams for $ub\to td$ process at NNLO QCD.}
 \label{fig:Feynman diagrams}
\end{figure}

As was stated in \cite{Assadsolimani:2014oga}, the color conservation would reduce the number of needed two-loop Feynman diagrams for non-factorizable corrections to the $t$-channel single-top production. It can be understood from the example (see Fig. \ref{fig:color conserve}) in which the interference of two-loop and tree diagram vanishes due to the presence of single $\text{SU}(3)$ generator in a trace. 
Eventually, only 18 diagrams will survive and 9 of them are depicted in Fig. \ref{fig:Feynman diagrams}. 
The remaining ones can be obtained by crossing the external legs from the 9 diagrams listed in Fig. \ref{fig:Feynman diagrams}. It turns out that they share the same master integrals as the former 9 diagrams up to different inputs, say $t \leftrightarrow u$ for example. 

The 9 Feynman diagrams shown in Fig. \ref{fig:Feynman diagrams} are categorized in two ways. Firstly, there are 3 topologies of the propagators, namely, the planar double box (db) diagrams, left-crossed double box (lxb) diagrams, and the right-crossed double box (rxb) diagrams. Each topology contains 3 diagrams. The diagrams with the same topology are arranged in the same row in Fig. \ref{fig:Feynman diagrams}. Secondly, the diagrams are categorized by the number of internal massive propagator(s). This number ranges from 1 to 3. The diagrams with the same number of internal massive propagator(s) are arranged in the same column in Fig. \ref{fig:Feynman diagrams}.
Among the two categorizing criteria, the topology and the number of massive internal propagator(s), the latter affects more on the properties of the integrals in the family. As introduced in section \ref{sec:review}, the canonical differential equation method is suitable to evaluate the master integrals in terms of transcendental functions like GPL. However, it is not always guaranteed that we can find a UT basis in a family, especially for those with too-many internal massive propagators. In these families, different kinds of square roots may appear in the attempt to transform the differential equations into a canonical form. Sometimes we cannot find a transformation of the kinematic variables to simultaneously rationalize all the square roots. The above situation may cause the difficulty brought by elliptic functions. In these cases, the solutions of the differential equations cannot be expressed in form of transcendental functions like GPL. In many cases, they are in form of elliptic functions as well as the iterative integration of elliptic functions \cite{Caron-Huot:2012awx,Bonciani:2016qxi,Primo:2016ebd}. These functions are more complicated and their properties and numerical evaluation are not as well-studied as GPL functions. 

As we will see later in this section, the diagrams with 1 internal massive propagator in the first column of Fig. \ref{fig:Feynman diagrams} are with good properties. The differential equations of 1-mass (with 1 internal massive propagator) planar and left-crossed double box are free of square roots. As for the 1-mass right-crossed diagram, there exists one square root inside and it can be rationalized after a redefinition of kinematic variables. For these diagrams, the canonical differential equations can be built, and the analytic results of the integrals can be expressed in GPL functions by solving the canonical differential equations using the method discussed in section \ref{sec:review}. The results of the 3 diagrams are shown in the next subsections. The construction of the differential equation in the rest 6 diagrams, with more internal masses, are more subtle due to the existence of more square roots and possible elliptic sectors. The maximal cut from Baikov representation \cite{Baikov:1996cd,Baikov:1996rk,Baikov:2005nv} also provides us some hints about the information of the underlining square roots and elliptic sectors. If a simultaneous rationalization of these square roots does not exist, corresponding elliptic sectors appear in these diagrams. In such a case, we are not aiming to express the corresponding integrals into iterative integrations of elliptic functions. However, the expansion of the kinematic variables provides us with another approach to get around elliptic functions. We are going to discuss this in subsection \ref{subsec: rest 6 diagrams}.

\subsection{The 1-mass planar double box diagram}
\label{subsec:db1}
\begin{figure}[h]
\centering
\includegraphics[width=0.45\textwidth]{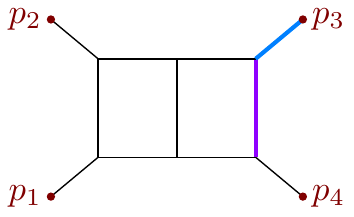}
\caption{The planar double box topology.}
\label{fig:db1}
\end{figure}

In this section, we present the analytic results of master integrals of the planar double box diagram with 1 internal mass. The diagram is shown in Fig. \ref{fig:db1}.

The propagators of the planar double box diagram are defined as
\begin{align}\label{eq:propagators-db1}
D_1&= l_1^2, &
D_2&= \left(l_1+p_1\right){}^2, &
D_3&= \left(l_1+p_1+p_2\right){}^2,\nonumber\\
D_4&= \left(l_2-p_1-p_2\right){}^2, &
D_5&= \left(l_2+p_4\right){}^2-m_W^2, &
D_6&= l_2^2,\nonumber\\
D_7&= \left(l_1+l_2\right){}^2, &
D_8&= \left(l_1-p_4\right){}^2, &
D_9&= \left(l_2-p_1\right){}^2.
\end{align}

Using FIRE6 \cite{Smirnov:2019qkx}, we found 31 master integrals for this diagram, which means we need to determine 31 UT integrals to build the canonical differential equations. In this family, the UT integrals we chose to form the basis are listed below.
\begin{align}\label{eq:UT db1 [[1]]}
I_1=& t G_{0,1,1,0,1,1,1,0,0}-m_t^2 G_{0,1,1,1,1,1,1,0,-1}-s G_{1,0,1,1,0,1,1,0,0}+m_t^2 G_{1,0,1,1,1,0,1,0,0}\nonumber\\&-s m_W^2 G_{1,0,1,1,1,1,1,0,0}+t G_{1,1,0,1,1,0,1,0,0}-s t G_{1,1,1,1,1,1,0,0,0}\nonumber\\&+s G_{1,1,1,1,1,1,1,-1,-1},\\
I_2=& -s \left(-t+m_t^2\right) G_{0,1,1,1,1,1,1,0,0}+s t G_{1,1,0,1,1,1,1,0,0}+s^2 G_{1,1,1,1,1,1,1,-1,0}\nonumber\\&-s^2 t G_{1,1,1,1,1,1,1,0,0},\\
I_3=& t G_{1,1,0,1,1,0,1,0,0}+s G_{1,1,1,1,1,0,1,-1,0}-s t G_{1,1,1,1,1,0,1,0,0},\\
I_4=& -\frac{s (-1+\epsilon ) m_W^2 G_{1,0,0,0,2,0,2,0,0}}{\epsilon ^3 \left(-m_t^2+m_W^2\right)}-\frac{6 m_t^2 \left(-s+m_t^2-m_W^2\right) G_{1,0,0,1,1,0,2,0,0}}{\epsilon  \left(m_t^2-m_W^2\right)}\nonumber\\&+\frac{s m_W^2 \left(s-m_t^2+m_W^2\right) G_{2,0,0,1,1,0,2,0,0}}{\epsilon ^2 \left(-m_t^2+m_W^2\right)},\\
I_5=& -t G_{0,1,1,0,1,1,1,0,0}+\left(-s+m_t^2\right) G_{0,1,1,1,1,1,1,0,-1},\\
I_6=& -s G_{0,1,1,0,1,1,1,0,0}+s G_{1,1,1,0,1,1,1,-1,0},\\
I_7=& -\frac{2 m_W^2 G_{0,0,1,0,2,0,2,0,0}}{\epsilon ^2}+\frac{\left(m_t^2-m_W^2\right) G_{0,0,2,0,1,0,2,0,0}}{\epsilon ^2},\\
I_8=& s^2 \left(-t+m_W^2\right) G_{1,1,1,1,1,1,1,0,0},\quad
I_9= -s \left(s-m_t^2\right) G_{1,1,1,1,1,1,1,0,-1},\\
I_{10}=& s \left(-t+m_t^2\right) G_{1,1,1,1,1,0,1,0,0},\quad
I_{11}= s \left(-t+m_W^2\right) G_{1,1,0,1,1,1,1,0,0},\\
I_{12}=& s \left(-s+m_t^2\right) G_{1,0,1,1,1,1,1,0,0},\quad
I_{13}= s \left(-t+m_W^2\right) G_{0,1,1,1,1,1,1,0,0},\\
I_{14}=& s^2 G_{1,1,1,1,0,1,1,0,0},\quad
I_{15}= s t G_{1,1,1,0,1,1,1,0,0},\\
I_{16}=& \left(-s-t+m_t^2\right) G_{0,1,1,0,1,1,1,0,0},\quad
I_{17}= \left(-s+m_t^2\right) G_{1,0,1,1,1,0,1,0,0},\\
I_{18}=& \left(-s+m_t^2\right) G_{1,0,1,0,1,1,1,0,0},\quad
I_{19}= \left(-t+m_t^2\right) G_{0,1,1,1,1,0,1,0,0},\\
I_{20}=& (s+t) G_{1,1,0,1,1,0,1,0,0},\quad
I_{21}= t G_{1,1,0,0,1,1,1,0,0},\\
I_{22}=& s G_{0,1,1,1,0,1,1,0,0},\quad
I_{23}= -\frac{\left(-s+m_t^2\right) G_{0,0,1,0,1,1,2,0,0}}{\epsilon },\\
I_{24}=& \frac{\left(-s+m_t^2\right) G_{1,0,1,0,2,0,1,0,0}}{\epsilon },\quad
I_{25}= \frac{\left(-s+m_t^2\right) G_{1,0,1,0,1,0,2,0,0}}{\epsilon },\\
I_{26}=& \frac{\left(-s+m_t^2\right) G_{1,0,0,1,1,0,2,0,0}}{\epsilon },\quad
I_{27}= -\frac{(-1+\epsilon ) m_W^2 G_{1,0,0,0,2,0,2,0,0}}{\epsilon ^3},\\
I_{28}=& \frac{s m_W^2 G_{1,0,2,0,3,0,0,0,0}}{\epsilon ^3},\quad
I_{29}= \frac{t G_{0,1,0,0,2,0,2,0,0}}{\epsilon ^2},\\
I_{30}=& \frac{s G_{0,0,2,0,0,2,1,0,0}}{\epsilon ^2},\quad
I_{31}= \frac{m_t^2 G_{0,0,1,0,2,0,2,0,0}}{\epsilon ^2}
\label{eq:UT db1 [[-1]]}
\end{align}

We are explaining how we determined this UT basis. At first, we used the package \texttt{DlogBasis} \cite{Henn:2020lye} to generate several $d\text{log}$ integrals. As was conjectured in \cite{Chen:2020uyk}, the $d\text{log}$ integrals should be UT integrals. Among the $d\text{log}$ integrals, we selected 20 linearly independent ones, to serve as part of the UT basis. They are $I_i$ for
\begin{equation}\label{eq:dlog db1}
    i\in\{1\sim3,{5,6},8\sim22\}.
\end{equation}

The master integrals of this diagram form a 31-dimensional linear space. Thus, the 20 $d\text{log}$ integrals do not form a complete basis. We need to choose additional 11 integrals to form a linearly complete UT basis. A better choice of the sectors that need to be completed is shown in Table \ref{tab: manual sectors db1}.
\begin{table}[h]
    \centering
    \begin{tabular}{|c|c|c|}
    \hline
         sector & label&\# of integrals \\
         \hline
         (1,0,1,0,1,0,0,0,0)&\{1,3,5\}&1 \\
         \hline
         (1,0,0,0,1,0,1,0,0)&\{1,5,7\}&1 \\
         \hline
         (0,1,0,0,1,0,1,0,0)&\{2,5,7\}&2 \\
         \hline
         (0,0,1,0,1,0,1,0,0)&\{3,5,7\}&2 \\
         \hline
         (0,0,1,0,0,1,1,0,0)&\{3,6,7\}&1 \\
         \hline
         (1,0,1,0,1,0,1,0,0)&\{1,3,5,7\}&2 \\
         \hline
         (1,0,0,1,1,0,1,0,0)&\{1,4,5,7\}&2 \\
         \hline
         (0,0,1,0,1,1,1,0,0)&\{3,5,6,7\}&1 \\
         \hline
         
    \end{tabular}
    \caption{Sectors to be completed of the double box diagram.}
    \label{tab: manual sectors db1}
\end{table}
These sectors are either sunset or bubble-triangle diagrams, we can easily determine a UT basis in each of the sectors. There are in total 12 UT integrals found in this step. Together with the $d\text{log}$ integrals in \eqref{eq:dlog db1}, there are 32 UT integrals. We chose 31 (linearly independent) of them by our preference to form the complete UT basis $I_1\sim I_{31}$ shown above.

With the UT basis determined, we derived the corresponding differential equations and the $\Tilde{A}$ matrix. Their results are in the attached files. See Appendix \ref{appendix: attached files} for details. The $\Tilde{A}$ matrix we derived are in form as \eqref{eq:Atilde in letters}, with symbol letters
\begin{align}
W_1&= m_t^2,\quad
W_2= m_W^2,\quad
W_3= s,\quad
W_4= t,\quad
W_5= s+t,\quad
W_6= m_t^2-m_W^2,\\
W_7&= -s+m_t^2,\quad
W_8= -t+m_t^2,\quad
W_9= -t+m_W^2,\quad
W_{10}= -s+m_t^2-m_W^2,\\
W_{11}&= -s-t+m_t^2,\quad
W_{12}= -t m_t^2+s m_W^2+t m_W^2.
\end{align}
With the canonical differential equation derived, we can perform an iterative integration on it to get the analytic results of the integrals, with some undetermined boundary constants. Then, we need to analyze the singularities of the UT integrals. Once a singularity point is found spurious, it generates a constraint on these undetermined constants. We can do this analysis using the methods mentioned in \ref{sec:solving DE}. We have found that the UT integrals $I_1,\cdots, I_{31}$ are all finite in the limits
\begin{equation}
    s \rightarrow -t, \quad s \rightarrow m_t^2-m_W^2, \quad s \rightarrow m_t^2, \quad m_t^2 \rightarrow 0.
\end{equation}
By requiring regularity of all the UT integrals at the above spurious singularity points, the undetermined constants of the boundaries can be completely fixed, up to one independent input integral which is easy to get a compact form
\begin{equation}
\begin{aligned}
    I_{28} &= \frac{\Gamma(1+\epsilon)^2}{\epsilon^4(m_W^2)^{2\epsilon}}(-\frac{s}{m_W^2})^{-\epsilon}\frac{\Gamma(1-\epsilon)^2}{2\Gamma(1-2\epsilon)}.
\end{aligned}
\end{equation}
Using these, the differential equations are solved completely. Notice that the differential equation is solved in the physical region $s>m_t^2>m_W^2>0, m_t^2-s < t <0$. This is the same for the next two families. 

The UT integrals are now expressed in the form of $\epsilon$-expansion with the coefficient being GPL functions, shown below.
\begin{align}\label{eq:analytic_UT_db1[[1]]}
&\bar{I}_{1}=\mathcal{O}(1),\quad
\bar{I}_{2}=\mathcal{O}(1),\quad
\bar{I}_{3}=\mathcal{O}(\frac{1}{\epsilon ^2}),\\&
\bar{I}_{4}=-\frac{1}{\epsilon ^4}+\frac{-2 i \pi +2 G(0,x)-2 G(1,z)}{\epsilon ^3}+\mathcal{O}(\frac{1}{\epsilon ^2}),\quad
\bar{I}_{5}=\mathcal{O}(\frac{1}{\epsilon }),\\&
\bar{I}_{6}=-\frac{G(-1,y)}{2 \epsilon ^3}+\mathcal{O}(\frac{1}{\epsilon ^2}),\quad
\bar{I}_{7}=\frac{1}{\epsilon ^4}-\frac{2 G(1,z)}{\epsilon ^3}+\mathcal{O}(\frac{1}{\epsilon ^2}),\\&
\bar{I}_{8}=-\frac{1}{4 \epsilon ^4}+\frac{-\frac{i \pi }{2}+G(-1,y)+\frac{1}{2} G(0,x)-\frac{1}{2} G(1,z)}{\epsilon ^3}+\mathcal{O}(\frac{1}{\epsilon ^2}),\quad
\bar{I}_{9}=\mathcal{O}(\frac{1}{\epsilon ^2}),\\&
\bar{I}_{10}=\frac{G(-1,y)-G(1,z)}{2 \epsilon ^3}+\mathcal{O}(\frac{1}{\epsilon ^2}),\quad
\bar{I}_{11}=\mathcal{O}(\frac{1}{\epsilon ^2}),\quad
\bar{I}_{12}=\mathcal{O}(1),\quad
\bar{I}_{13}=\mathcal{O}(\frac{1}{\epsilon ^2}),\\&
\bar{I}_{14}=\frac{1}{4 \epsilon ^4}+\frac{\frac{i \pi }{2}-\frac{1}{2} G(0,x)}{\epsilon ^3}+\mathcal{O}(\frac{1}{\epsilon ^2}),\quad
\bar{I}_{15}=-\frac{G(-1,y)}{2 \epsilon ^3}+\mathcal{O}(\frac{1}{\epsilon ^2}),\quad
\bar{I}_{16}=\mathcal{O}(\frac{1}{\epsilon }),\\&
\bar{I}_{17}=\mathcal{O}(1),\quad
\bar{I}_{18}=\mathcal{O}(1),\quad
\bar{I}_{19}=\mathcal{O}(\frac{1}{\epsilon ^2}),\quad
\bar{I}_{20}=\mathcal{O}(\frac{1}{\epsilon }),\quad
\bar{I}_{21}=\mathcal{O}(\frac{1}{\epsilon ^2}),\\&
\bar{I}_{22}=\mathcal{O}(\frac{1}{\epsilon ^2}),\quad
\bar{I}_{23}=\mathcal{O}(\frac{1}{\epsilon ^2}),\quad
\bar{I}_{24}=\mathcal{O}(\frac{1}{\epsilon }),\quad
\bar{I}_{25}=\mathcal{O}(\frac{1}{\epsilon ^2}),\quad
\bar{I}_{26}=\mathcal{O}(\frac{1}{\epsilon ^2}),\\&
\bar{I}_{27}=\frac{1}{\epsilon ^4}+\mathcal{O}(\frac{1}{\epsilon ^2}),\quad
\bar{I}_{28}=\frac{1}{2 \epsilon ^4}+\frac{\frac{i \pi }{2}-\frac{1}{2} G(0,x)}{\epsilon ^3}+\mathcal{O}(\frac{1}{\epsilon ^2}),\\&
\bar{I}_{29}=-\frac{G(-1,y)}{\epsilon ^3}+\mathcal{O}(\frac{1}{\epsilon ^2}),\quad
\bar{I}_{30}=\frac{1}{\epsilon ^4}+\frac{2 i \pi -2 G(0,x)}{\epsilon ^3}+\mathcal{O}(\frac{1}{\epsilon ^2}),\\&
\bar{I}_{31}=-\frac{G(1,z)}{\epsilon ^3}+\mathcal{O}(\frac{1}{\epsilon ^2}),\label{eq:analytic_UT_db1[[-1]]}
\end{align}

 where we have defined scaleless variables as 
 \begin{equation}\label{eq:scaleless variables xyz}
     x=\frac{s}{m_W^2},\;
     y=-\frac{t}{m_W^2},\;
     z=\frac{m_t^2}{m_W^2}.
 \end{equation}
 Note that, in the results \eqref{eq:analytic_UT_db1[[1]]}$\sim$\eqref{eq:analytic_UT_db1[[-1]]}, we have multiplied the UT integrals with a regulator as 
 \begin{equation}\label{eq:I bar regulator}
    \bar{I}_i=\frac{(m_W^2)^{2 \epsilon}}{\Gamma(1+\epsilon)^2} I_i,
\end{equation}
in order to keep the Euler gamma constant $\gamma_E$ absent in the $\epsilon$ expansion and to keep the UT integrals dimensionless. The results shown in \eqref{eq:analytic_UT_db1[[1]]}$\sim$\eqref{eq:analytic_UT_db1[[-1]]} only kept the expressions up to $\mathcal{O}(\frac{1}{\epsilon ^2})$, considering the next orders are with too-long expressions. We put the results that are kept up to order $\mathcal{O}(1)$ in the attached file "db1/analytic\_UT.txt", which is a Mathematica readable list of the analytic expressions of $\{\bar{I}_1,\cdots,\bar{I}_{31}\}$. See appendix \ref{appendix: attached files} for more details about the attached files.

We performed numerical checks on the results shown in \eqref{eq:analytic_UT_db1[[1]]}$\sim$\eqref{eq:analytic_UT_db1[[-1]]}. The numerical validation point is 
\begin{equation}\label{eq:numeric check point}
    s = 360000, \;t = -\frac{235765}{2}, \;m_W^2 = 6400, \;m_t^2=29929.
\end{equation}
We evaluated \eqref{eq:analytic_UT_db1[[1]]}$\sim$\eqref{eq:analytic_UT_db1[[-1]]} on this check point using GiNaC \cite{Bauer:2000cp,Vollinga:2004sn,Vollinga:2005pk,Weinzierl:2002hv}\footnote{Recently, there is another C++ package developed for fast GPL evaluation, named {\sc FastGPL} \cite{Wang:2021imw}.}. 
 We label the numeric results as $(\bar{I}_i)_{\text{GPL}}$. Besides, we directly computed the $\epsilon$ expansion of the Laporta Master integrals using AMFlow \cite{Liu:2022chg}, and using \eqref{eq:UT db1 [[1]]}$\sim$\eqref{eq:UT db1 [[-1]]} to convert it to numeric expressions of the UT integrals. We label the numerical UT integrals derived from this way by $(\bar{I}_i)_{\text{AMF}}$.

 The analytic expressions of the UT integrals in this family passed the numerical check with\footnote{The precision goal in AMFlow is set to be 30 reliable digits.} 
\begin{equation}
    |(\bar{I}_i)_{\text{GPL}}-(\bar{I}_i)_{\text{AMF}}|<10^{-26}
\end{equation}
at each order from $\epsilon^{-4}$ to $\epsilon^{0}$ on the numeric point \eqref{eq:numeric check point}. The numerical results of $(\bar{I}_i)_{\text{GPL}}$ and $(\bar{I}_i)_{\text{AMF}}$ on this check point are put in the attached files "db1/numericUT\_GPL.m" and "db1/numericUT\_AMF.m", respectively. Some selected results are digested in Table \ref{tab:numeric db1}.
\begin{table}[htbp]
    \centering
    \begin{tabular}{|c|c|l|}
    \hline
    integral & $\epsilon$ order  & \multicolumn{1}{|c|}{coefficient}\\
    \hline
    $(\bar{I}_4)_{\text{GPL}}$ & $\epsilon^{-4}$ &-1.00000000000000000000000000000\\
    \hline
    $(\bar{I}_4)_{\text{AMF}}$ & $\epsilon^{-4}$ &-1.00000000000000000000000000000\\
    \hline
    $(\bar{I}_4)_{\text{GPL}}$ & $\epsilon^{-3}$ &5.45574065710303313100762912852\\
    \hline
    $(\bar{I}_4)_{\text{AMF}}$ & $\epsilon^{-3}$ &5.45574065710303313100762912852\\
    \hline
    $(\bar{I}_8)_{\text{GPL}}$ & $\epsilon^{-4}$ &-0.250000000000000000000000000000\\
    \hline
    $(\bar{I}_8)_{\text{AMF}}$ & $\epsilon^{-4}$ &-0.250000000000000000000000000000\\
    \hline
    $(\bar{I}_8)_{\text{GPL}}$ & $\epsilon^{-3}$ &4.33019437410005361889679452697\\
    \hline
    $(\bar{I}_8)_{\text{AMF}}$ & $\epsilon^{-3}$ &4.33019437410005361889679452697\\
    \hline
    \end{tabular}
    \caption{Selected numeric results in the planar double box family.}
    \label{tab:numeric db1}
\end{table}

\subsection{The 1-mass left-crossed double box diagram}
\label{subsec:lxb1}
\begin{figure}[h]
\centering
\includegraphics[width=0.45\textwidth]{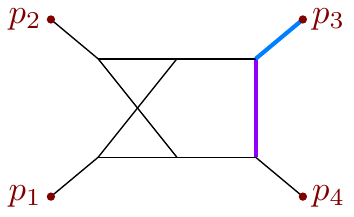}
\caption{The left-crossed double box topology.}
\label{fig:lxb1}
\end{figure}

In this section, we present the analytic results of master integrals of the left-crossed double box diagram with 1 internal mass. The diagram is shown in Fig. \ref{fig:lxb1}.
The propagators of the left-crossed double box diagram are defined as
\begin{align}\label{eq:propagators-lxb1}
D_1&= l_1^2,&
D_2&= \left(l_1+p_1\right){}^2,&
D_3&= \left(l_2-p_1-p_2\right){}^2,\nonumber\\
D_4&= \left(l_2+p_4\right){}^2-m_W^2,&
D_5&= l_2^2,&
D_6&= \left(l_1+l_2\right){}^2,\nonumber\\
D_7&= \left(l_1+l_2-p_2\right){}^2,&
D_8&= \left(l_1-p_4\right){}^2,&
D_9&= \left(l_2-p_1\right){}^2.
\end{align}

Using FIRE6 \cite{Smirnov:2019qkx}, we found 35 master integrals in this diagram. We can find their combinations to form a UT basis which contains 35 UT integrals. The UT basis we found in this family is as follows.

\begin{align}
I_1=& -s t G_{0,1,1,1,1,1,1,0,0}+s m_t^2 G_{0,1,1,1,1,1,1,0,0}-s m_W^2 G_{1,0,1,1,1,1,1,0,0}\nonumber\\&+s t G_{1,1,1,1,0,1,1,0,0}-s m_t^2 G_{1,1,1,1,0,1,1,0,0}-s t G_{1,1,1,1,1,0,1,0,0}\nonumber\\&+s t G_{1,1,1,1,1,1,0,0,0}-s^2 G_{1,1,1,1,1,1,1,-1,0}+s m_W^2 G_{1,1,1,1,1,1,1,0,-1}\nonumber\\&+s^2 m_W^2 G_{1,1,1,1,1,1,1,0,0},\\
I_2=& s m_t^2 G_{0,1,1,1,1,1,1,0,0}-s m_W^2 G_{0,1,1,1,1,1,1,0,0}-s m_W^2 G_{1,0,1,1,1,1,1,0,0}\nonumber\\&-s t G_{1,1,0,1,1,1,1,0,0}-s t G_{1,1,1,1,1,0,1,0,0}+s t G_{1,1,1,1,1,1,0,0,0}\nonumber\\&-s^2 G_{1,1,1,1,1,1,1,-1,0}+s t G_{1,1,1,1,1,1,1,0,-1}+s^2 t G_{1,1,1,1,1,1,1,0,0},\\
I_3=& -s G_{0,1,1,1,1,1,1,0,-1}+s G_{1,0,1,1,0,1,1,0,0}-s m_W^2 G_{1,0,1,1,1,1,1,0,0}\nonumber\\&-s G_{1,1,1,1,0,1,1,-1,0}+s t G_{1,1,1,1,0,1,1,0,0}-s t G_{1,1,1,1,1,0,1,0,0}\nonumber\\&+s G_{1,1,1,1,1,1,1,-1,-1},\\
I_4=& s G_{1,0,1,1,0,1,1,0,0}-t G_{1,1,0,1,0,1,1,0,0}+s G_{1,1,1,1,0,0,1,0,0}+t G_{1,1,1,1,0,1,0,0,0}\nonumber\\&-s G_{1,1,1,1,0,1,1,-1,0}+t G_{1,1,1,1,0,1,1,0,-1}+s t G_{1,1,1,1,0,1,1,0,0},\\
I_5=& \frac{s (1+\epsilon ) \left(s+t-m_t^2\right) G_{0,1,0,1,1,1,2,0,0}}{\epsilon ^2}+\frac{s \left(2 m_t^2+m_W^2\right) G_{0,1,0,2,0,0,2,0,0}}{\epsilon ^2 \left(-t+m_t^2\right)}\nonumber\\&-\frac{s \left(2 t+m_W^2\right) G_{0,1,0,2,0,2,0,0,0}}{\epsilon ^2 \left(-t+m_t^2\right)}-\frac{s \left(m_t^2-m_W^2\right) G_{0,2,0,1,0,0,2,0,0}}{2 \epsilon ^2 \left(-t+m_t^2\right)}\nonumber\\&+\frac{s \left(t-m_W^2\right) G_{0,2,0,1,0,2,0,0,0}}{2 \epsilon ^2 \left(-t+m_t^2\right)},\\
I_6=& -t G_{1,0,0,1,1,1,1,0,0}-s G_{1,0,1,1,1,1,1,0,-1}+m_t^2 G_{1,0,1,1,1,1,1,0,-1}\nonumber\\&+s t G_{1,0,1,1,1,1,1,0,0}+s m_W^2 G_{1,0,1,1,1,1,1,0,0},\\
I_7=& -t G_{1,1,-1,1,1,1,1,0,0}+t G_{1,1,0,1,1,1,1,0,-1}+s t G_{1,1,0,1,1,1,1,0,0},\\
I_8=& -t G_{0,1,0,1,1,1,1,0,0}-s G_{0,1,1,1,1,1,1,0,-1}+m_t^2 G_{0,1,1,1,1,1,1,0,-1},\\
I_9=& -s G_{0,1,0,1,1,1,1,0,0}+s G_{1,1,0,1,1,1,1,-1,0},\\
I_{10}=& \frac{3 m_W^2 G_{1,0,1,1,0,2,0,0,0}}{\epsilon }+\frac{m_W^2 \left(s-m_t^2+m_W^2\right) G_{1,0,1,2,0,2,0,0,0}}{\epsilon ^2},\\
I_{11}=& -\frac{2 m_W^2 G_{0,1,0,2,0,0,2,0,0}}{\epsilon ^2}+\frac{\left(m_t^2-m_W^2\right) G_{0,2,0,1,0,0,2,0,0}}{\epsilon ^2},\\
I_{12}=& -s^2 G_{1,1,1,1,1,0,1,0,0}-s t G_{1,1,1,1,1,0,1,0,0}+s m_t^2 G_{1,1,1,1,1,0,1,0,0}\nonumber\\&-s m_W^2 G_{1,1,1,1,1,0,1,0,0},\\
I_{13}=& -s^2 G_{1,0,1,1,1,1,1,0,0}-s t G_{1,0,1,1,1,1,1,0,0}+s m_t^2 G_{1,0,1,1,1,1,1,0,0}\nonumber\\&-s m_W^2 G_{1,0,1,1,1,1,1,0,0},\\
I_{14}=& -s t G_{1,1,1,1,1,1,0,0,0}+s m_W^2 G_{1,1,1,1,1,1,0,0,0},\\
I_{15}=& s m_t^2 G_{1,1,1,1,0,1,1,0,0}-s m_W^2 G_{1,1,1,1,0,1,1,0,0},\\
I_{16}=& -s t G_{0,1,1,1,1,1,1,0,0}+s m_W^2 G_{0,1,1,1,1,1,1,0,0},\\
I_{17}=& s m_W^2 G_{1,1,0,1,1,1,1,0,0},\quad
I_{18}= s G_{1,1,1,0,1,1,1,0,-1},\\
I_{19}=& -s G_{1,0,0,1,1,1,1,0,0}-t G_{1,0,0,1,1,1,1,0,0}+m_t^2 G_{1,0,0,1,1,1,1,0,0},\\
I_{20}=& -t G_{1,0,1,1,0,1,1,0,0}+m_t^2 G_{1,0,1,1,0,1,1,0,0},\\
I_{21}=& -t G_{0,1,1,1,0,1,1,0,0}+m_t^2 G_{0,1,1,1,0,1,1,0,0},\\
I_{22}=& s G_{1,1,1,1,0,1,0,0,0}+t G_{1,1,1,1,0,1,0,0,0},\\
I_{23}=& s G_{1,1,1,1,0,0,1,0,0}+t G_{1,1,1,1,0,0,1,0,0},\\
I_{24}=& \left(s+t-m_t^2\right) G_{0,1,0,1,1,1,1,0,0},\quad
I_{25}= t G_{1,1,0,1,1,1,0,0,0},\\
I_{26}=& t G_{1,1,0,1,1,0,1,0,0},\quad
I_{27}= s G_{1,1,0,1,0,1,1,0,0},\quad
I_{28}= s G_{1,1,0,0,1,1,1,0,0},\\
I_{29}=& s G_{0,1,1,0,1,1,1,0,0},\quad
I_{30}= \frac{\left(-s+m_t^2\right) G_{1,0,1,1,0,2,0,0,0}}{\epsilon },\\
I_{31}=& \frac{\left(-s+m_t^2\right) G_{0,1,0,1,1,0,2,0,0}}{\epsilon },\quad
I_{32}= -\frac{(-1+\epsilon ) m_W^2 G_{2,0,0,1,0,2,0,0,0}}{\epsilon ^2 (1+\epsilon )},\\
I_{33}=& -\frac{\left(-s-t+m_t^2\right) G_{2,0,0,2,0,0,1,0,0}}{\epsilon ^2},\quad
I_{34}= \frac{t G_{0,1,0,2,0,2,0,0,0}}{\epsilon ^2},\\
I_{35}=& \frac{m_t^2 G_{0,1,0,2,0,0,2,0,0}}{\epsilon ^2}.
\end{align}

The method we use to find the above UT basis is similar to those introduced in the former subsection. Among the 35 UT integrals found, 25 of them were found by \texttt{DlogBasis} \cite{Henn:2020lye}. They are $I_i$ for
\begin{equation}\label{eq:dlog lxb1}
    i\in\{1\sim 4,6\sim 9,12\sim 19,21\sim29\}.
\end{equation}

Then, we found UT integrals in the following sectors manually, shown in Table \ref{tab: manual sectors lxb1}.
\begin{table}[h]
    \centering
    \begin{tabular}{|c|c|c|}
    \hline
         sector & label&\# of integrals \\
         \hline
         (1,0,0,1,0,0,1,0,0)&\{1,4,7\}&2 \\
         \hline
         (0,1,0,1,0,1,0,0,0)&\{2,4,6\}&2 \\
         \hline
         (0,1,0,1,0,0,1,0,0)&\{2,4,7\}&2 \\
         \hline
         (1,0,1,1,0,1,0,0,0)&\{1,3,4,6\}&2 \\
         \hline
         (0,1,0,1,1,0,1,0,0)&\{2,4,5,7\}&1 \\
         \hline
         (0,1,0,1,1,1,1,0,0)&\{2,4,5,6,7\}&2 \\
         \hline
         
    \end{tabular}
    \caption{Sectors to be completed of the left-crossed box diagram}
    \label{tab: manual sectors lxb1}
\end{table}
There are in total 11 integrals in Table \ref{tab: manual sectors lxb1}. Together with the 25 integrals found by \texttt{DlogBasis}, we can select (by some preference) 35 independent UT integrals to form a complete 35-dimensional UT basis. 

The canonical differential equation and matrix $\Tilde{A}$ can be found in the attached files. Also, see Appendix \ref{appendix: attached files} for details. The symbol letters that appear in $\Tilde{A}$ are
\begin{align}
W_1&= m_t^2,\quad
W_2= m_W^2,\quad
W_3= s,\quad
W_4= t,\quad
W_5= s+t,\quad
W_6= m_t^2-m_W^2,\\
W_7&= -s+m_t^2,\quad
W_8= -t+m_t^2,\quad
W_9= -t+m_W^2,\quad
W_{10}= -s+m_t^2-m_W^2,\\
W_{11}&= -s-t+m_t^2,\quad
W_{12}= -t m_t^2+s m_W^2+t m_W^2,\quad
W_{13}= -s-t+m_t^2-m_W^2,\\
W_{14}&= -s t-t^2+t m_t^2+s m_W^2,\quad
W_{15}= -s m_t^2-t m_t^2+m_t^4+t m_W^2-m_t^2 m_W^2.
\end{align}

Similar to the last section, we found the UT integrals $I_1,\cdots,I_{35}$ are all regular when
\begin{equation}
    s \rightarrow -t, \quad s \rightarrow m_t^2-m_W^2, \quad t \rightarrow 0, \quad t \rightarrow m_t^2.
\end{equation}
Constrains followed by these spurious singularities fix all the boundary constants up to one simple input integral,
\begin{equation}
\begin{aligned}
    I_{32} &= \frac{\Gamma(1+\epsilon)^2}{\epsilon^4(m_W^2)^{2\epsilon}}\frac{-\Gamma(1-\epsilon)\Gamma(1+2\epsilon)}{\Gamma(1+\epsilon)}.
\end{aligned}
\end{equation}
With these, the canonical differential equations are solved and the results are put in the attached file "lxb1/analytic\_UT.txt". Notice that, same as the last section, the content of this file is a list as $\{\bar{I}_1,\cdots,\bar{I}_{35}\}$, where $\bar{I}_i$ is still what was defined in \eqref{eq:I bar regulator}. The results in the files were kept up to $\mathcal{O}(1)$. We digested them here, up to $\mathcal{O}(\frac{1}{\epsilon^2})$, as follows
\begin{small}
\begin{align}
&\bar{I}_{1}=-\frac{3}{4 \epsilon ^4}+\frac{-\frac{3 i \pi }{2}+\frac{3}{2} G(0,x)-\frac{1}{2} G(1,z)-G(1-z,y)-G(y+z-1,x)}{\epsilon ^3}+\mathcal{O}(\frac{1}{\epsilon ^2}),\\&
\bar{I}_{2}=\mathcal{O}(\frac{1}{\epsilon ^2}),\quad
\bar{I}_{3}=\mathcal{O}(\frac{1}{\epsilon ^2}),\quad
\bar{I}_{4}=\mathcal{O}(\frac{1}{\epsilon ^2}),\quad
\bar{I}_{5}=\frac{G(-1,y)-G(1,z)}{\epsilon ^3}+\mathcal{O}(\frac{1}{\epsilon ^2}),\\&
\bar{I}_{6}=\mathcal{O}(\frac{1}{\epsilon ^2}),\quad
\bar{I}_{7}=-\frac{G(-1,y)}{2 \epsilon ^3}+\mathcal{O}(\frac{1}{\epsilon ^2}),\quad
\bar{I}_{8}=\mathcal{O}(\frac{1}{\epsilon }),\\&
\bar{I}_{9}=-\frac{G(-1,y)}{2 \epsilon ^3}+\mathcal{O}(\frac{1}{\epsilon ^2}),\quad
\bar{I}_{10}=\frac{-i \pi +G(0,x)-G(1,z)}{\epsilon ^3}+\mathcal{O}(\frac{1}{\epsilon ^2}),\\&
\bar{I}_{11}=\frac{1}{\epsilon ^4}-\frac{2 G(1,z)}{\epsilon ^3}+\mathcal{O}(\frac{1}{\epsilon ^2}),\quad
\bar{I}_{12}=\mathcal{O}(\frac{1}{\epsilon ^2}),\quad
\bar{I}_{13}=\mathcal{O}(\frac{1}{\epsilon ^2}),\quad
\bar{I}_{14}=\mathcal{O}(\frac{1}{\epsilon ^2}),\\&
\bar{I}_{15}=\frac{1}{4 \epsilon ^4}\!+\!\frac{\frac{i \pi }{2}\!+\!\frac{1}{2} G(-1,y)\!-\!\frac{1}{2} G(0,x)\!-\!\frac{1}{2} G(1,z)\!+\!\frac{1}{2} G(1\!-\!z,y)\!+\!\frac{1}{2} G(y\!+\!z\!-\!1,x)}{\epsilon ^3}\!+\!\mathcal{O}(\frac{1}{\epsilon ^2}),\!\\&
\bar{I}_{16}=\mathcal{O}(\frac{1}{\epsilon ^2}),\\&
\bar{I}_{17}=\frac{-1}{4 \epsilon ^4}\!+\!\frac{-\frac{i \pi }{2}\!-\!\frac{1}{2} G(-1,y)\!+\!\frac{1}{2} G(0,x)\!-\!\frac{1}{2} G(1,z)\!-\!\frac{1}{2} G(1\!-\!z,y)\!-\!\frac{1}{2} G(y\!+\!z\!-\!1,x)}{\epsilon ^3}\!+\!\mathcal{O}(\frac{1}{\epsilon ^2}),\\&
\bar{I}_{18}=-\frac{1}{4 \epsilon ^4}+\frac{-\frac{i \pi }{2}+\frac{1}{2} G(0,x)}{\epsilon ^3}+\mathcal{O}(\frac{1}{\epsilon ^2}),\quad
\bar{I}_{19}=\mathcal{O}(\frac{1}{\epsilon ^2}),\quad
\bar{I}_{20}=\mathcal{O}(\frac{1}{\epsilon }),\quad
\bar{I}_{21}=\mathcal{O}(\frac{1}{\epsilon ^2}),\\&
\bar{I}_{22}=\mathcal{O}(\frac{1}{\epsilon }),\quad
\bar{I}_{23}=\mathcal{O}(\frac{1}{\epsilon ^2}),\quad
\bar{I}_{24}=\mathcal{O}(\frac{1}{\epsilon }),\quad
\bar{I}_{25}=\mathcal{O}(\frac{1}{\epsilon ^2}),\quad
\bar{I}_{26}=\mathcal{O}(\frac{1}{\epsilon }),\quad
\bar{I}_{27}=\mathcal{O}(\frac{1}{\epsilon ^2}),\\&
\bar{I}_{28}=\frac{1}{4 \epsilon ^4}+\frac{\frac{i \pi }{2}-\frac{1}{2} G(0,x)}{\epsilon ^3}+\mathcal{O}(\frac{1}{\epsilon ^2}),\quad
\bar{I}_{29}=\mathcal{O}(\frac{1}{\epsilon ^2}),\quad
\bar{I}_{30}=\mathcal{O}(\frac{1}{\epsilon ^2}),\quad
\bar{I}_{31}=\mathcal{O}(\frac{1}{\epsilon ^2}),\\&
\bar{I}_{32}=-\frac{1}{\epsilon ^4}+\mathcal{O}(\frac{1}{\epsilon ^2}),\quad
\bar{I}_{33}=\frac{G(1,z)+G(1-z,y)+G(-1+y+z,x)}{\epsilon ^3}+\mathcal{O}(\frac{1}{\epsilon ^2}),\\&
\bar{I}_{34}=-\frac{G(-1,y)}{\epsilon ^3}+\mathcal{O}(\frac{1}{\epsilon ^2}),\quad
\bar{I}_{35}=-\frac{G(1,z)}{\epsilon ^3}+\mathcal{O}(\frac{1}{\epsilon ^2})
\end{align}
\end{small}
where the variables $x$, $y$, and $z$ are of the same definition as in \eqref{eq:scaleless variables xyz}.

We did the same numerical check on point \eqref{eq:numeric check point} and got the result of $(\bar{I}_i)_{\text{GPL}}$ and $(\bar{I}_i)_{\text{AMF}}$ of this family. They are put in attached files "lxb1/numericUT\_GPL.m" and "lxb1/numericUT\_AMF.m", respectively. The results passed the numeric check with
\begin{equation}
    |(\bar{I}_i)_{\text{GPL}}-(\bar{I}_i)_{\text{AMF}}|<10^{-25}
\end{equation}
at each order from $\epsilon^{-4}$ to $\epsilon^{0}$. Some selected numerical results are shown in Table \ref{tab:numeric lxb1}.
\begin{table}[htbp]
    \centering
    \begin{tabular}{|c|c|c|l|}
    \hline
    integral & $\epsilon$ order & part  & \multicolumn{1}{|c|}{coefficient}\\
    \hline
    $(\bar{I}_1)_{\text{GPL}}$ & $\epsilon^{-4}$ &real &-0.750000000000000000000000000000\\
    \hline
    $(\bar{I}_1)_{\text{AMF}}$ & $\epsilon^{-4}$ &real &-0.750000000000000000000000000000\\
    \hline
    $(\bar{I}_1)_{\text{GPL}}$ & $\epsilon^{-3}$ &real &3.16478394140259919434483588045\\
    \hline
    $(\bar{I}_1)_{\text{AMF}}$ & $\epsilon^{-3}$ &real &3.16478394140259919434483588045\\
    \hline
    $(\bar{I}_1)_{\text{GPL}}$ & $\epsilon^{-3}$ &imaginary & -6.28318530717958647692528676656\\
    \hline
    $(\bar{I}_1)_{\text{AMF}}$ & $\epsilon^{-3}$ &imaginary & -6.28318530717958647692528676656\\
    \hline
    
    \end{tabular}
    \caption{Selected numeric results in the planar left-crossed box family.}
    \label{tab:numeric lxb1}
\end{table}

\subsection{The 1-mass right-crossed double box diagram}
\label{subsec:rxb1}

\begin{figure}[h]
\centering
\includegraphics[width=0.45\textwidth]{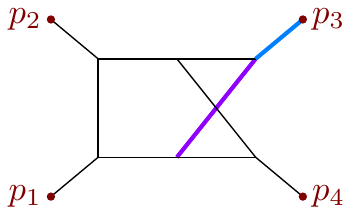}
\caption{The right-crossed double box topology.}
\label{fig:rxb1}
\end{figure}
In this subsection, we present the UT integrals and their corresponding analytic $\epsilon$ expansion results of the right-crossed double box diagram with 1 internal mass (Fig. \ref{fig:rxb1}). In this diagram, the propagators are defined as
\begin{align}\label{eq:propagators-rxb1}
D_1&= l_1^2,&
D_2&= \left(l_1+p_1\right){}^2,&
D_3&= \left(l_1+p_1+p_2\right){}^2,\nonumber\\
D_4&= \left(l_2+p_4\right){}^2,&
D_5&= l_2^2,&
D_6&= \left(l_1+l_2\right){}^2-m_W^2,\nonumber\\
D_7&= \left(l_1+l_2+p_1+p_2+p_4\right){}^2,&
D_8&= \left(l_1-p_4\right){}^2,&
D_9&= \left(l_2-p_1\right){}^2.
\end{align}
The number of the master integrals in this family is 55, determined using FIRE6 \cite{Smirnov:2019qkx}. Thus, we need to find 55 UT integrals to form a linearly independent and complete basis. The method that we found the UT integrals in this diagram is different from what we used in the last sections. The reason is that we encounter a square root in this family. As a consequence, we found it difficult to generate enough $d\text{log}$ integrals using \texttt{DlogBasis}. For the top sector, we cannot generate a $d\text{log}$ integral using this method. Thus, in this section, we used the Magnus series to construct a UT basis.

At first, we need to determine a linear basis. There is no general way, according to our knowledge, to find such a linear basis. It is always obtained in an empirical way through trial and error. We first prepare a set of master integrals defined below.
\begin{align}
T_{1} &= G_{1,1,1,1,1,1,1,0,-1}, &
T_{2} &= G_{1,1,1,1,1,1,1,-1,0}, &
T_{3} &= G_{1,1,1,1,1,1,1,0,0}, \nn
T_{4} &= G_{0,1,1,1,1,1,1,0,-1}, &
T_{5} &= G_{0,1,1,1,1,1,1,0,0}, &
T_{6} &= G_{1,0,1,1,1,1,1,0,0}, \nn
T_{7} &= \frac{G_{1,1,0,1,1,2,1,0,0}}{\epsilon }, &
T_{8} &= G_{1,1,0,1,1,1,1,-1,0}, &
T_{9} &= G_{1,1,0,1,1,1,1,0,0}, \nn
T_{10} &= G_{1,1,1,0,1,1,1,-1,0}, &
T_{11} &= \frac{G_{1,1,1,0,1,2,1,0,0}}{\epsilon }, &
T_{12} &= G_{1,1,1,0,1,1,1,0,0}, \nn
T_{13} &= G_{1,1,1,1,0,1,1,0,0}, &
T_{14} &= \frac{G_{0,2,0,1,1,1,1,0,0}}{\epsilon }, &
T_{15} &= G_{0,1,0,1,1,1,1,0,0}, \nn
T_{16} &= G_{1,0,0,1,1,1,1,0,0}, &
T_{17} &= \frac{G_{0,1,1,0,2,1,1,0,0}}{\epsilon }, &
T_{18} &= G_{0,1,1,0,1,1,1,0,0}, \nn
T_{19} &= \frac{G_{1,0,1,0,1,1,2,0,0}}{\epsilon }, &
T_{20} &= \frac{G_{1,0,2,0,1,1,1,0,0}}{\epsilon }, &
T_{21} &= G_{1,0,1,0,1,1,1,0,0}, \nn
T_{22} &= G_{1,1,0,0,1,1,1,0,0}, &
T_{23} &= G_{1,0,1,1,0,1,1,0,0}, &
T_{24} &= \frac{G_{1,1,0,2,0,1,1,0,0}}{\epsilon }, \nn
T_{25} &= G_{1,1,0,1,0,1,1,0,0}, &
T_{26} &= G_{1,1,0,1,1,0,1,0,0}, &
T_{27} &= \frac{G_{1,1,1,0,1,0,2,0,0}}{\epsilon }, \nn
T_{28} &= \frac{G_{0,1,1,1,1,2,0,0,0}}{\epsilon }, &
T_{29} &= G_{0,1,1,1,1,1,0,0,0}, &
T_{30} &= G_{1,0,1,1,1,1,0,0,0}, \nn
T_{31} &= G_{1,1,0,1,1,1,0,0,0}, &
T_{32} &= \frac{G_{1,1,1,1,0,3,0,0,0}}{\epsilon ^2}, &
T_{33} &= \frac{G_{1,1,1,1,0,2,0,0,0}}{\epsilon }, \nn
T_{34} &= \frac{G_{0,0,1,0,2,1,1,0,0}}{\epsilon }, &
T_{35} &= \frac{G_{0,1,0,0,2,2,1,0,0}}{\epsilon ^2}, &
T_{36} &= \frac{G_{0,1,0,0,2,1,1,0,0}}{\epsilon }, \nn
T_{37} &= \frac{(1-2 \epsilon ) G_{2,0,0,0,1,1,1,0,0}}{\epsilon ^2},&
T_{38} &= \frac{G_{0,2,0,1,0,1,1,0,0}}{\epsilon }, &
T_{39} &= \frac{G_{2,0,0,1,0,2,1,0,0}}{\epsilon ^2}, \nn
T_{40} &= \frac{G_{2,0,0,1,0,1,1,0,0}}{\epsilon }, &
T_{41} &= \frac{G_{1,0,2,0,0,2,1,0,0}}{\epsilon ^2}, &
T_{42} &= \frac{G_{1,0,1,0,1,0,2,0,0}}{\epsilon }, \nn
T_{43} &= \frac{G_{1,0,1,1,0,3,0,0,0}}{\epsilon ^2}, &
T_{44} &= \frac{G_{1,0,1,1,0,2,0,0,0}}{\epsilon }, &
T_{45} &= \frac{G_{0,1,0,0,2,0,2,0,0}}{\epsilon ^2}, \nn
T_{46} &= \frac{G_{1,0,0,0,2,0,2,0,0}}{\epsilon ^2}, &
T_{47} &= \frac{G_{1,0,0,2,0,0,2,0,0}}{\epsilon ^2}, &
T_{48} &= \frac{G_{0,0,2,0,2,1,0,0,0}}{\epsilon ^2}, \nn
T_{49} &= \frac{G_{0,0,1,0,2,2,0,0,0}}{\epsilon ^2}, &
T_{50} &= \frac{G_{0,0,2,2,0,1,0,0,0}}{\epsilon ^2}, &
T_{51} &= \frac{G_{0,0,1,2,0,2,0,0,0}}{\epsilon ^2}, \nn
T_{52} &= \frac{G_{0,2,0,2,0,1,0,0,0}}{\epsilon ^2}, &
T_{53} &= \frac{G_{0,1,0,2,0,2,0,0,0}}{\epsilon ^2}, &
T_{54} &= \frac{(1-\epsilon ) G_{1,0,0,2,0,2,0,0,0}}{\epsilon ^3}, \nn
T_{55} &= \frac{G_{1,0,2,0,0,2,0,0,0}}{\epsilon ^2}. &
\end{align}
The corresponding differential equation is almost linear in $\epsilon$ except for the homogeneous part of sector $(1,1,0,1,1,1,1,0,0)$. It has the following form,
\begin{equation}
\begin{pmatrix}
1+\epsilon & 1+\epsilon & \frac{1+\epsilon+\epsilon^2}{\epsilon} \\
\epsilon & 1+\epsilon & 1+\epsilon \\
\epsilon & \epsilon & 1+\epsilon
\end{pmatrix},
\end{equation}
where the entries only present the characters in $\epsilon$, For instance, $\epsilon$ means the factorization of the dimensional regulator. To transform the final piece into linear form, we first notice that the leading singularity of the integral $T_{9}$ is
\begin{equation}
    T_9|_{\text{Leading Singularity}} = \sqrt{t^2 (m_W^2)^2+ \left[4 s t (s+t)-2 t m_t^2 (2 s+t)\right]m_W^2+t^2 (m_t^2)^2} \equiv r.
\end{equation}
This means $rT_9$ should be a UT integral. With it in mind, we make an {\it ansatz} that the new master integrals (in this sector) defined by
\begin{equation}
\begin{pmatrix}
    T_{7}^{\prime} \\
    T_{8}^{\prime} \\
    T_{9}^{\prime}
\end{pmatrix}
=
    \begin{pmatrix}
        1 & 0 & \chi/\epsilon \\
        0 & 1 & 0 \\
        0 & 0 & r
    \end{pmatrix}
\begin{pmatrix}
    T_{7} \\
    T_{8} \\
    T_{9}
\end{pmatrix}
\end{equation}
would produce a linear form. Meeting such a requirement sets up a first-order differential equation of the unknown function $\chi$. The solution reads
\begin{equation}
    \chi = \frac{2 s \left(s+t-m_t^2\right)+t \left(m_W^2-m_t^2\right)}{4 s m_W^2 \left(s-m_t^2\right)+t \left[4 s m_W^2+\left(m_t^2-m_W^2\right){}^2\right]}.
\end{equation}
Now the linear basis is ready, then one can use the Magnux series method to obtain a UT one. In our practice, the Magnus series \eqref{eq:Magnus series} terminates at $n=2$, i.e. $\Omega_i=0$ for $i>2$. The UT integrals given by the Magnus series method are as follows.
\begin{align}
I_{1}=&-\frac{3 t G_{0,1,0,0,2,0,2,0,0}}{4 \epsilon ^2}+\frac{3 t m_t^2 G_{1,0,0,0,2,0,2,0,0}}{4 \epsilon ^2 \left(-s-t+m_t^2\right)}+\frac{3 s t G_{1,0,0,2,0,0,2,0,0}}{4 \epsilon ^2 \left(s+t-m_t^2\right)}\nonumber\\&-\frac{3 t^2 G_{1,1,0,1,1,0,1,0,0}}{-s-t+m_t^2}-\frac{1}{2} t \left(-2 s+m_t^2-m_W^2\right) G_{1,1,0,1,1,1,1,0,0}\nonumber\\&-\frac{s t G_{1,1,1,0,1,0,2,0,0}}{\epsilon }-s \left(-s+m_t^2\right) G_{1,1,1,1,1,1,1,-1,0}\nonumber\\&-s \left(-s+m_t^2\right) G_{1,1,1,1,1,1,1,0,-1}+s \left(-s+m_t^2\right) \left(t+m_W^2\right) G_{1,1,1,1,1,1,1,0,0},\\
I_{2}=&-\frac{1}{2} t \left(m_t^2-m_W^2\right) G_{1,1,0,1,1,1,1,0,0}-s \left(-s-t+m_t^2\right) G_{1,1,1,1,1,1,1,-1,0}\nonumber\\&+s \left(-s-t+m_t^2\right) m_W^2 G_{1,1,1,1,1,1,1,0,0},\\
I_{3}=&-\left(-t+m_t^2\right) \left(-s+m_t^2-m_W^2\right) G_{0,1,1,1,1,1,1,0,0}\nonumber\\&+t \left(-s+m_t^2-m_W^2\right) G_{1,1,0,1,1,1,1,0,0}-s \left(s-m_t^2+m_W^2\right) G_{1,1,1,1,1,1,1,-1,0}\nonumber\\&+s t \left(s-m_t^2+m_W^2\right) G_{1,1,1,1,1,1,1,0,0},\\
I_{4}=&-\frac{3 t \left(-s+m_t^2\right) G_{0,0,1,0,2,1,1,0,0}}{4 \epsilon  \left(-s-t+m_t^2\right)}-\frac{3 t m_W^2 G_{0,0,1,0,2,2,0,0,0}}{8 \epsilon ^2 \left(-s-t+m_t^2\right)}\nonumber\\&+\frac{3 t \left(s-m_W^2\right) G_{0,0,2,0,2,1,0,0,0}}{16 \epsilon ^2 \left(-s-t+m_t^2\right)}-\frac{t G_{0,1,0,0,2,0,2,0,0}}{8 \epsilon ^2}-\frac{3 t \left(-s+m_t^2\right) G_{0,1,0,1,1,1,1,0,0}}{4 \left(-s-t+m_t^2\right)}\nonumber\\&+\frac{3 t m_W^2 G_{0,1,0,2,0,2,0,0,0}}{8 \epsilon ^2 \left(-s-t+m_t^2\right)}-\frac{\left(4 s^2+4 s t+3 t^2-8 s m_t^2-7 t m_t^2+4 m_t^4\right) G_{0,1,1,0,1,1,1,0,0}}{4 \left(-s-t+m_t^2\right)}\nonumber\\&+\frac{t \left(s-m_W^2\right) G_{0,1,1,0,2,1,1,0,0}}{8 \epsilon }+\frac{1}{2} t G_{0,1,1,1,1,1,0,0,0}+\left(-s+m_t^2\right) G_{0,1,1,1,1,1,1,0,-1}\nonumber\\&-\frac{1}{2} t \left(-2 s+m_t^2+m_W^2\right) G_{0,1,1,1,1,1,1,0,0}+\frac{3 t \left(-t+m_t^2\right) G_{0,2,0,1,0,1,1,0,0}}{4 \epsilon  \left(-s-t+m_t^2\right)}\nonumber\\&-\frac{t \left(t-m_W^2\right) G_{0,2,0,1,1,1,1,0,0}}{8 \epsilon }-\frac{3 t \left(t-m_W^2\right) G_{0,2,0,2,0,1,0,0,0}}{16 \epsilon ^2 \left(-s-t+m_t^2\right)},\\
I_{5}=&\left(-s-t+m_t^2\right) \left(m_t^2-m_W^2\right) G_{0,1,1,1,1,1,1,0,0},\\
I_{6}=&\left(-s+m_t^2\right) \left(-s+m_t^2-m_W^2\right) G_{1,0,1,1,1,1,1,0,0},\\
I_{7}=&\frac{m_t^2 \left(-s-t+m_t^2\right) G_{0,1,0,0,2,0,2,0,0}}{\epsilon ^2 \left(m_t^2+m_W^2\right)}-\frac{4 (s+t) m_t^2 G_{0,1,0,0,2,1,1,0,0}}{\epsilon  \left(m_t^2+m_W^2\right)}\nonumber\\&+\frac{8 (s+t) m_t^2 G_{0,1,0,1,1,1,1,0,0}}{m_t^2+m_W^2}\nonumber\\&-\frac{\left(-s-t+m_t^2\right) \left(-2 t m_t^2+2 s m_W^2+m_t^2 m_W^2+m_W^4\right) G_{0,2,0,1,1,1,1,0,0}}{\epsilon  \left(m_t^2+m_W^2\right)}\nonumber\\&-\frac{m_t^4 G_{1,0,0,0,2,0,2,0,0}}{\epsilon ^2 \left(m_t^2+m_W^2\right)}-\frac{4 (s+t) m_t^2 G_{1,1,0,1,0,1,1,0,0}}{m_t^2+m_W^2}-\frac{8 (s+t) m_t^2 G_{1,1,0,1,1,1,1,-1,0}}{m_t^2+m_W^2}\nonumber\\&+\Big((2 t \epsilon  m_t^4+2 s^2 m_W^2+2 s t m_W^2+8 s^2 \epsilon  m_W^2+8 s t \epsilon  m_W^2-2 s m_t^2 m_W^2-t m_t^2 m_W^2\nonumber\\&\quad-8 s \epsilon  m_t^2 m_W^2+t m_W^4+6 t \epsilon  m_W^4 )G_{1,1,0,1,1,1,1,0,0}\Big)/{\epsilon  \left(m_t^2+m_W^2\right)}\nonumber\\&+\frac{m_W^2 \left(t m_t^4+4 s^2 m_W^2+4 s t m_W^2-4 s m_t^2 m_W^2-2 t m_t^2 m_W^2+t m_W^4\right) G_{1,1,0,1,1,2,1,0,0}}{\epsilon  \left(m_t^2+m_W^2\right)}\nonumber\\&+\frac{s \left(-2 t m_t^2+2 s m_W^2+m_t^2 m_W^2+m_W^4\right) G_{1,1,0,2,0,1,1,0,0}}{\epsilon  \left(m_t^2+m_W^2\right)},\\
I_{8}=&\left(-t-m_t^2\right) G_{0,1,0,1,1,1,1,0,0}+(s+t) G_{1,1,0,1,1,1,1,-1,0}\nonumber\\&-\frac{1}{2} t \left(m_t^2+m_W^2\right) G_{1,1,0,1,1,1,1,0,0},\\
I_{9}=&r G_{1,1,0,1,1,1,1,0,0},\\
I_{10}=&t G_{1,1,0,0,1,1,1,0,0}+s G_{1,1,1,0,1,1,1,-1,0}-s t G_{1,1,1,0,1,1,1,0,0},\\
I_{11}=&-2 s m_W^2 G_{1,1,1,0,1,1,1,0,0}-\frac{s m_W^2 \left(-s-t+m_W^2\right) G_{1,1,1,0,1,2,1,0,0}}{\epsilon },\\
I_{12}=&s (s+t) G_{1,1,1,0,1,1,1,0,0},\quad
I_{13}=s \left(-t+m_t^2\right) G_{1,1,1,1,0,1,1,0,0},\\
I_{14}=&-\frac{\left(s+t-m_t^2\right) \left(-t+m_W^2\right) G_{0,2,0,1,1,1,1,0,0}}{\epsilon },\\
I_{15}=&\left(-s+m_t^2\right) G_{0,1,0,1,1,1,1,0,0},\\
I_{16}=&\left(-s+m_t^2\right) G_{1,0,0,1,1,1,1,0,0},\quad
I_{17}=-\frac{\left(s+t-m_t^2\right) \left(-s+m_W^2\right) G_{0,1,1,0,2,1,1,0,0}}{\epsilon },\\
I_{18}=&\left(-t+m_t^2\right) G_{0,1,1,0,1,1,1,0,0},\quad
I_{19}=\frac{\left(-s+m_t^2\right) \left(m_t^2-m_W^2\right) G_{1,0,1,0,1,1,2,0,0}}{\epsilon },\\
I_{20}=&\frac{s \left(s-m_t^2\right) G_{1,0,2,0,1,1,1,0,0}}{\epsilon },\quad
I_{21}=\left(-s+m_t^2\right) G_{1,0,1,0,1,1,1,0,0},\\
I_{22}=&(s+t) G_{1,1,0,0,1,1,1,0,0},\quad
I_{23}=\left(-s+m_t^2\right) G_{1,0,1,1,0,1,1,0,0},\\
I_{24}=&\frac{s \left(-t+m_W^2\right) G_{1,1,0,2,0,1,1,0,0}}{\epsilon },\quad
I_{25}=(s+t) G_{1,1,0,1,0,1,1,0,0},\\
I_{26}=&t G_{1,1,0,1,1,0,1,0,0},\quad
I_{27}=\frac{s \left(s+t-m_t^2\right) G_{1,1,1,0,1,0,2,0,0}}{\epsilon },\\
I_{28}=&\frac{\left(s t-s m_W^2-t m_W^2+m_t^2 m_W^2\right) G_{0,1,1,1,1,2,0,0,0}}{\epsilon },\\
I_{29}=&\left(-s-t+m_t^2\right) G_{0,1,1,1,1,1,0,0,0},\quad
I_{30}=\left(-s+m_t^2\right) G_{1,0,1,1,1,1,0,0,0},\\
I_{31}=&t G_{1,1,0,1,1,1,0,0,0},\\
I_{32}=&\frac{s m_W^2 G_{1,1,1,1,0,2,0,0,0}}{\epsilon }+\frac{s m_W^2 \left(-t+m_W^2\right) G_{1,1,1,1,0,3,0,0,0}}{\epsilon ^2},\\
I_{33}=&\frac{s t G_{1,1,1,1,0,2,0,0,0}}{\epsilon },\quad
I_{34}=\frac{\left(-s+m_t^2\right) G_{0,0,1,0,2,1,1,0,0}}{\epsilon },\\
I_{35}=&\frac{3 m_W^2 G_{0,1,0,0,2,1,1,0,0}}{\epsilon }+\frac{m_W^2 \left(-s-t+m_W^2\right) G_{0,1,0,0,2,2,1,0,0}}{\epsilon ^2},\\
I_{36}=&\frac{(s+t) G_{0,1,0,0,2,1,1,0,0}}{\epsilon },\\
I_{37}=&\frac{m_t^4 G_{1,0,0,0,2,0,2,0,0}}{2 \epsilon ^2 \left(m_t^2-m_W^2\right)}+\frac{(-1+\epsilon ) m_t^2 m_W^2 G_{1,0,0,2,0,2,0,0,0}}{2 \epsilon ^3 \left(m_t^2-m_W^2\right)}\nonumber\\&+\frac{(-1+2 \epsilon ) m_t^2 m_W^2 G_{2,0,0,0,1,1,1,0,0}}{\epsilon ^2 \left(m_t^2-m_W^2\right)},\\
I_{38}=&\frac{\left(-t+m_t^2\right) G_{0,2,0,1,0,1,1,0,0}}{\epsilon },\\
I_{39}=&-\frac{3 m_W^2 G_{2,0,0,1,0,1,1,0,0}}{\epsilon }-\frac{m_W^2 \left(s-m_t^2+m_W^2\right) G_{2,0,0,1,0,2,1,0,0}}{\epsilon ^2},\\
I_{40}=&\frac{\left(-s+m_t^2\right) G_{2,0,0,1,0,1,1,0,0}}{\epsilon },\quad
I_{41}=\frac{s m_t^2 G_{1,0,2,0,0,2,1,0,0}}{\epsilon ^2},\\
I_{42}=&\frac{\left(-s+m_t^2\right) G_{1,0,1,0,1,0,2,0,0}}{\epsilon },\quad
I_{43}=\frac{\left(-s+m_t^2\right) m_W^2 G_{1,0,1,1,0,3,0,0,0}}{\epsilon ^2},\\
I_{44}=&\frac{\left(-s+m_t^2\right) G_{1,0,1,1,0,2,0,0,0}}{\epsilon },\quad
I_{45}=\frac{\left(-s-t+m_t^2\right) G_{0,1,0,0,2,0,2,0,0}}{\epsilon ^2},\\
I_{46}=&\frac{m_t^2 G_{1,0,0,0,2,0,2,0,0}}{\epsilon ^2},\quad
I_{47}=\frac{s G_{1,0,0,2,0,0,2,0,0}}{\epsilon ^2},\\
I_{48}=&\frac{2 m_W^2 G_{0,0,1,0,2,2,0,0,0}}{\epsilon ^2}+\frac{\left(-s+m_W^2\right) G_{0,0,2,0,2,1,0,0,0}}{\epsilon ^2},\quad
I_{49}=\frac{s G_{0,0,1,0,2,2,0,0,0}}{\epsilon ^2},\\
I_{50}=&-\frac{2 m_W^2 G_{0,0,1,2,0,2,0,0,0}}{\epsilon ^2}+\frac{\left(m_t^2-m_W^2\right) G_{0,0,2,2,0,1,0,0,0}}{\epsilon ^2},\quad
I_{51}=\frac{m_t^2 G_{0,0,1,2,0,2,0,0,0}}{\epsilon ^2},\\
I_{52}=&\frac{2 m_W^2 G_{0,1,0,2,0,2,0,0,0}}{\epsilon ^2}+\frac{\left(-t+m_W^2\right) G_{0,2,0,2,0,1,0,0,0}}{\epsilon ^2},\quad
I_{53}=\frac{t G_{0,1,0,2,0,2,0,0,0}}{\epsilon ^2},\\
I_{54}=&-\frac{(-1+\epsilon ) m_W^2 G_{1,0,0,2,0,2,0,0,0}}{\epsilon ^3},\quad
I_{55}=\frac{s G_{1,0,2,0,0,2,0,0,0}}{\epsilon ^2}.
\end{align}

The matrix $\Tilde{A}$ can be found in the attached files (see Appendix \ref{appendix: attached files}). The symbol letters for this diagram that appear in $\Tilde{A}$ are
\begin{align}
W_1&= m_t^2,\quad
W_2= m_W^2,\quad
W_3= s,\quad
W_4= t,\quad
W_5= s+t,\quad
W_6= m_t^2-m_W^2,\\
W_7&= -s+m_t^2,\quad
W_8= -s+m_W^2,\quad
W_9= -t+m_t^2,\quad
W_{10}= -t+m_W^2,\\
W_{11}&= -s+m_t^2-m_W^2,\quad
W_{12}= -s-t+m_t^2,\quad
W_{13}= -s-t+m_W^2,\\
W_{14}&= -t m_t^2+(s+t) m_W^2,\quad
W_{15}= -s m_W^2+m_t^2 \left(-t+m_W^2\right),\\
W_{16}&= -t m_W^2+m_t^2 \left(-s+m_W^2\right),\quad
W_{17}= s t-(s+t) m_W^2+m_t^2 m_W^2,\\
W_{18}&= 4 s \left(s-m_t^2\right) m_W^2+t \left(4 s m_W^2+\left(m_t^2-m_W^2\right){}^2\right),\\
W_{19}&= \frac{{r+f_{19}}}{{r}-f_{19}},\quad
W_{20}= \frac{{\left(s+m_W^2\right)r+f_{20}}}{{\left(s+m_W^2\right)r}-f_{20}},\quad
W_{21}= \frac{{\left(s+t-m_W^2\right)r+f_{21}}}{{\left(s+t-m_W^2\right)r}-f_{21}},
\end{align}
where
\begin{align}
    f_{19}=&t m_t^2-(2 s+t) m_W^2,\\
    f_{20}=&s t m_t^2+2 s^2 m_W^2+3 s t m_W^2-2 s m_t^2 m_W^2-t m_t^2 m_W^2+t m_W^4,\\
    f_{21}=&s t m_t^2+t^2 m_t^2+2 s^2 m_W^2+s t m_W^2-t^2 m_W^2-2 s m_t^2 m_W^2-t m_t^2 m_W^2+t m_W^4,
\end{align}

Considering there is a square root $r$ in the differential equations, before solving them, we need to rationalize it. This can be achieved by the following change of variables
\begin{equation}
    x=x_1, y=\frac{(x_1+y_1)^2}{x_1-y_1^2-y_1 z_1}, z=z_1+1
\end{equation}
where the variables $x$, $y$, and $z$ are those defined in \eqref{eq:scaleless variables xyz}. After this, the differential equations become rational. To solve the canonical differential equations, we used the following spurious singularities to fix the boundary constants,
\begin{itemize}
    \item $I_{15,18,28,29,49}$ are finite at $s\to 0$.
    \item $I_{10,11,25,27,36}$ are finite at $s\to -t$.
    \item $I_{6,40,43}$ are finite at $s\to m_t^2-m_W^2$.
    \item $I_{13,25,28,29,32,33}$ are finite at $s\to m_t^2-t$.
    \item $I_{1,2,5,6,15,19,20,21,23,26,28,29,34,40,42,43,44}$ are finite at $s\to m_t^2$.
    \item $I_{18}$ is finite at $s\to \frac{\left(m_t^2-t\right) m_W^2}{m_t^2}$.
    \item $I_{1,3,5,7,8,10,11,12,25,26,27,53}$ are finite at $t\to 0$.
    \item $I_{22,30,31}$ each vanishes when $s=-t$, $s=m_t^2$, $t=0$, respectively.
\end{itemize}
Besides, we need the following compact results of some relatively simple UT integrals as input
\begin{equation}
    \begin{aligned}
        I_{41} &= \frac{\Gamma(1+\epsilon)^2}{\epsilon^4(m_W^2)^{2\epsilon}}(-\frac{s}{m_W^2})^{-\epsilon}\frac{\epsilon\Gamma(1-\epsilon)^3}{\Gamma(1-2\epsilon)\Gamma(2-\epsilon)}\frac{m_t^2}{m_W^2}{}_2F_1(1,1+\epsilon,2-\epsilon,\frac{m_t^2}{m_W^2}), \\
        I_{47} &= \frac{\Gamma(1+\epsilon)^2}{\epsilon^4(m_W^2)^{2\epsilon}}(-\frac{s}{m_W^2})^{-2\epsilon}\frac{\Gamma (1-\epsilon )^3 \Gamma (1+2 \epsilon)}{\Gamma (1-3 \epsilon ) \Gamma (1+\epsilon)^2},\\
        I_{55} &= \frac{\Gamma(1+\epsilon)^2}{\epsilon^4(m_W^2)^{2\epsilon}}(-\frac{s}{m_W^2})^{-\epsilon}\frac{-\Gamma(1-\epsilon)^2}{\Gamma(1-2\epsilon)},
    \end{aligned}
\end{equation}
where the expansion of the hypergeometric function is performed by using the HypExp package \cite{Huber:2005yg,Huber:2007dx}. With the conditions above, we are now able to solve the canonical differential equations. It is worthwhile to notice that not all the integrals have the dependence on square root $r$ in their differential equation. We solve the whole system with a bottom-up approach. Namely, we start from the lower sectors where no rationalization is needed at all. Then the higher sectors which depend on $r$ are included. Therefore the final results are a hybrid of GPLs with two distinct sets of variables.

After solving the differential equations, we get the analytic expressions of the UT basis. They are put in "rxb1/analytic\_UT.txt". Also, the content of this file is a list, whose members are analytic expressions of UT integrals multiplied by a regulator, as $\{\bar{I}_1,\cdots,\bar{I}_{55}\}$, defined in \eqref{eq:I bar regulator}. The results in the files were kept up to $\mathcal{O}(1)$. We digested them here and kept them up to $\mathcal{O}(\frac{1}{\epsilon^2})$, as follows 

\begin{align}
&\bar{I}_{1}=-\frac{3}{4 \epsilon ^4}+\frac{-\frac{i \pi }{2}+3 G\left(-1,y_1\right)+\cdots-\frac{3}{2} G\left(y_1^2,x_1\right)}{\epsilon ^3}+\mathcal{O}(\frac{1}{\epsilon ^2}),\\&
\bar{I}_{2}=-\frac{3}{4 \epsilon ^4}+\frac{\frac{3 i \pi }{2}+3 G\left(-1,y_1\right)+\cdots+\frac{3}{2} G\left(\frac{x_1 \left(1+x_1+2 y_1\right)}{y_1},z_1\right)}{\epsilon ^3}+\mathcal{O}(\frac{1}{\epsilon ^2}),\\&
\bar{I}_{3}=\mathcal{O}(\frac{1}{\epsilon ^2}),\quad
\bar{I}_{4}=\mathcal{O}(\frac{1}{\epsilon ^2}),\\&
\bar{I}_{5}=\frac{1}{4 \epsilon ^4}+\frac{\frac{i \pi }{2}+\frac{1}{2} G(-1,y)-\cdots-\frac{1}{2} G(y+z,x)}{\epsilon ^3}+\mathcal{O}(\frac{1}{\epsilon ^2}),\quad
\bar{I}_{6}=\mathcal{O}(\frac{1}{\epsilon ^2}),\\&
\bar{I}_{7}=\frac{2 G\left(-1,y_1\right)-\cdots-G\left(\frac{x_1 \left(1+x_1+2 y_1\right)}{y_1},z_1\right)}{\epsilon ^3}+\mathcal{O}(\frac{1}{\epsilon ^2}),\quad
\bar{I}_{8}=\mathcal{O}(\frac{1}{\epsilon ^2}),\\&
\bar{I}_{9}=\mathcal{O}(\frac{1}{\epsilon ^2}),\quad
\bar{I}_{10}=\mathcal{O}(\frac{1}{\epsilon }),\\&
\bar{I}_{11}=\frac{-i \pi +G(0,z)-G(1,z)+G(-z,y)+G(y+z,x)}{\epsilon ^3}+\mathcal{O}(\frac{1}{\epsilon ^2}),\quad
\bar{I}_{12}=\mathcal{O}(\frac{1}{\epsilon ^2}),\\&
\bar{I}_{13}=\frac{G(-1,y)-G(1,z)}{2 \epsilon ^3}+\mathcal{O}(\frac{1}{\epsilon ^2}),\\&
\bar{I}_{14}=\frac{1}{2 \epsilon ^4}\!+\!\frac{i \pi \!-\!2 G(-1,y)\!-\!G(0,z)\!+\!G(1,z)\!-\!G(-z,y)\!-\!G(y+z,x)}{\epsilon ^3}\!+\!\mathcal{O}(\frac{1}{\epsilon ^2}),\\&
\bar{I}_{15}=\mathcal{O}(\frac{1}{\epsilon }),\quad
\bar{I}_{16}=\mathcal{O}(\frac{1}{\epsilon }),\\&
\bar{I}_{17}=\frac{1}{2 \epsilon ^4}+\frac{i \pi -G(0,z)-2 G(1,x)+G(1,z)-G(-z,y)-G(y+z,x)}{\epsilon ^3}+\mathcal{O}(\frac{1}{\epsilon ^2}),\\&
\bar{I}_{18}=\mathcal{O}(\frac{1}{\epsilon }),\quad
\bar{I}_{19}=\frac{-G(0,x)+G(0,z)}{2 \epsilon ^3}+\mathcal{O}(\frac{1}{\epsilon ^2}),\\&
\bar{I}_{20}=\frac{G(1,x)-G(1,z)}{2 \epsilon ^3}+\mathcal{O}(\frac{1}{\epsilon ^2}),\quad
\bar{I}_{21}=\mathcal{O}(1),\quad
\bar{I}_{22}=\mathcal{O}(\frac{1}{\epsilon }),\quad
\bar{I}_{23}=\mathcal{O}(1),\\&
\bar{I}_{24}=\frac{1}{2 \epsilon ^4}+\frac{i \pi -2 G(-1,y)-G(0,x)+G(1,z)}{\epsilon ^3}+\mathcal{O}(\frac{1}{\epsilon ^2}),\quad
\bar{I}_{25}=\mathcal{O}(\frac{1}{\epsilon }),\\&
\bar{I}_{26}=\mathcal{O}(\frac{1}{\epsilon ^2}),\\&
\bar{I}_{27}=\frac{3}{2 \epsilon ^4}+\frac{3 i \pi -\frac{3}{2} G(0,x)-\frac{3}{2} G(0,z)-3 G(-z,y)-3 G(y+z,x)}{\epsilon ^3}+\mathcal{O}(\frac{1}{\epsilon ^2}),\\&
\bar{I}_{28}=\frac{G(-1,y)+G(1,x)-G(1,z)}{\epsilon ^3}+\mathcal{O}(\frac{1}{\epsilon ^2}),\quad
\bar{I}_{29}=\mathcal{O}(\frac{1}{\epsilon ^2}),\quad
\bar{I}_{30}=\mathcal{O}(\frac{1}{\epsilon }),\\&
\bar{I}_{31}=\mathcal{O}(\frac{1}{\epsilon ^2}),\quad
\bar{I}_{32}=\frac{1}{2 \epsilon ^4}+\frac{\frac{i \pi }{2}-G(-1,y)-\frac{1}{2} G(0,x)+\frac{1}{2} G(1,z)}{\epsilon ^3}+\mathcal{O}(\frac{1}{\epsilon ^2}),\\&
\bar{I}_{33}=\frac{G(-1,y)}{\epsilon ^3}+\mathcal{O}(\frac{1}{\epsilon ^2}),\quad
\bar{I}_{34}=\frac{-G(1,x)+G(1,z)}{2 \epsilon ^3}+\mathcal{O}(\frac{1}{\epsilon ^2}),\\&
\bar{I}_{35}=\frac{-i \pi +G(0,z)-G(1,z)+G(-z,y)+G(y+z,x)}{\epsilon ^3}+\mathcal{O}(\frac{1}{\epsilon ^2}),\quad
\bar{I}_{36}=\mathcal{O}(\frac{1}{\epsilon ^2}),\\&
\bar{I}_{37}=-\frac{G(1,z)}{\epsilon ^3}+\mathcal{O}(\frac{1}{\epsilon ^2}),\quad
\bar{I}_{38}=\frac{-G(-1,y)+G(1,z)}{2 \epsilon ^3}+\mathcal{O}(\frac{1}{\epsilon ^2}),\\&
\bar{I}_{39}=\frac{i \pi -G(0,x)+G(1,z)}{\epsilon ^3}+\mathcal{O}(\frac{1}{\epsilon ^2}),\quad
\bar{I}_{40}=\mathcal{O}(\frac{1}{\epsilon ^2}),\quad
\bar{I}_{41}=-\frac{G(1,z)}{\epsilon ^3}+\mathcal{O}(\frac{1}{\epsilon ^2}),\\&
\bar{I}_{42}=\frac{-G(0,x)+G(0,z)}{2 \epsilon ^3}+\mathcal{O}(\frac{1}{\epsilon ^2}),\quad
\bar{I}_{43}=\mathcal{O}(\frac{1}{\epsilon ^2}),\quad
\bar{I}_{44}=\mathcal{O}(\frac{1}{\epsilon }),\\&
\bar{I}_{45}=\frac{1}{\epsilon ^4}+\frac{2 i \pi -2 G(0,z)-2 G(-z,y)-2 G(y+z,x)}{\epsilon ^3}+\mathcal{O}(\frac{1}{\epsilon ^2}),\\&
\bar{I}_{46}=\frac{1}{\epsilon ^4}+\frac{2 i \pi -2 G(0,z)}{\epsilon ^3}+\mathcal{O}(\frac{1}{\epsilon ^2}),\quad
\bar{I}_{47}=\frac{1}{\epsilon ^4}+\frac{2 i \pi -2 G(0,x)}{\epsilon ^3}+\mathcal{O}(\frac{1}{\epsilon ^2}),\\&
\bar{I}_{48}=-\frac{1}{\epsilon ^4}+\frac{2 G(1,x)}{\epsilon ^3}+\mathcal{O}(\frac{1}{\epsilon ^2}),\quad
\bar{I}_{49}=-\frac{G(1,x)}{\epsilon ^3}+\mathcal{O}(\frac{1}{\epsilon ^2}),\\&
\bar{I}_{50}=\frac{1}{\epsilon ^4}-\frac{2 G(1,z)}{\epsilon ^3}+\mathcal{O}(\frac{1}{\epsilon ^2}),\quad
\bar{I}_{51}=-\frac{G(1,z)}{\epsilon ^3}+\mathcal{O}(\frac{1}{\epsilon ^2}),\\&
\bar{I}_{52}=-\frac{1}{\epsilon ^4}+\frac{2 G(-1,y)}{\epsilon ^3}+\mathcal{O}(\frac{1}{\epsilon ^2}),\quad
\bar{I}_{53}=-\frac{G(-1,y)}{\epsilon ^3}+\mathcal{O}(\frac{1}{\epsilon ^2}),\\&
\bar{I}_{54}=\frac{1}{\epsilon ^4}+\mathcal{O}(\frac{1}{\epsilon ^2}),\quad
\bar{I}_{55}=-\frac{1}{\epsilon ^4}+\frac{-i \pi +G(0,x)}{\epsilon ^3}+\mathcal{O}(\frac{1}{\epsilon ^2})
\end{align}

We performed numerical check on point \eqref{eq:numeric check point} and got the result of $(\bar{I}_i)_{\text{GPL}}$ and $(\bar{I}_i)_{\text{AMF}}$ of this family. Their results are put in attached files "rxb1/numericUT\_GPL.m" and "rxb1/numericUT\_AMF.m", respectively. The results passed the numeric check with
\begin{equation}
    |(\bar{I}_i)_{\text{GPL}}-(\bar{I}_i)_{\text{AMF}}|<10^{-30}
\end{equation}
at each order from $\epsilon^{-4}$ to $\epsilon^{0}$. Some selected numerical results are shown in Table \ref{tab:numeric rxb1}.
\begin{table}[htbp]
    \centering
    \begin{tabular}{|c|c|c|l|}
    \hline
    integral & $\epsilon$ order & part  & \multicolumn{1}{|c|}{coefficient}\\
    \hline
    $(\bar{I}_7)_{\text{GPL}}$ & $\epsilon^{-1}$ &real &219.075133281247447672103350945\\
    \hline
    $(\bar{I}_7)_{\text{AMF}}$ & $\epsilon^{-1}$ &real &219.075133281247447672103350945\\
    \hline
    $(\bar{I}_7)_{\text{GPL}}$ & $\epsilon^{-1}$ &imaginary &128.292541065870999789073454960 \\
    \hline
    $(\bar{I}_7)_{\text{AMF}}$ & $\epsilon^{-1}$ &imaginary &128.292541065870999789073454960 \\
    \hline
    $(\bar{I}_7)_{\text{GPL}}$ & $\epsilon^{0}$ &real &-473.853668694661084686278060658\\
    \hline
    $(\bar{I}_7)_{\text{AMF}}$ & $\epsilon^{0}$ &real &-473.853668694661084686278060658\\
    \hline
    $(\bar{I}_7)_{\text{GPL}}$ & $\epsilon^{0}$ &imaginary &471.438825850996829424630945814 \\
    \hline
    $(\bar{I}_7)_{\text{AMF}}$ & $\epsilon^{0}$ &imaginary &471.438825850996829424630945814 \\
    \hline

    \end{tabular}
    \caption{Selected numeric results in the non-planar right-crossed box family.}
    \label{tab:numeric rxb1}
\end{table}

\subsection{Discussions about the diagrams with multiple internal masses}\label{subsec: rest 6 diagrams}
In this subsection, we give a general discussion on the rest diagrams with more internal masses, which are not fully evaluated in this work. The diagrams in the second and the third columns in Fig. \ref{fig:Feynman diagrams} have more than one internal massive propagator. In these families, different complicated square roots would appear in the attempt of finding a canonical differential equation and they are not always simultaneously rationalizable. Thus, we find that it is not always possible to find a complete set of UT basis for each family to obtain a canonical differential equation. Even one can still construct UT integrals for the lower sector in these families, considering some of them may involve many square roots which can not be rationalized simultaneously, it is difficult to get a solution in terms of GPL. The main problem with these families is the appearance of elliptic sectors. Computing the maximal cut of an integral will hint at whether it is possible to express the integral in terms of GPL or if more complicated functions, such as elliptic integrals, will come into play \cite{Caron-Huot:2012awx,Bonciani:2016qxi}. In fact, the maximal cut provides the homogeneous solution of the corresponding differential equation \cite{Primo:2016ebd}. That may indicate the cut and uncut integrals, what we really need, live in the space of the same class of functions. As an example, we show here the maximal cut of two integrals in the family defined by the up-right corner diagram in Fig. \ref{fig:Feynman diagrams}. It is convenient to calculate the maximal cut of an integral in the Baikov representation. Using the loop-by-loop approach \cite{Frellesvig:2017aai}, we have
\begin{equation}
\begin{aligned}
\vcenter{\hbox{\includegraphics[scale = 1]{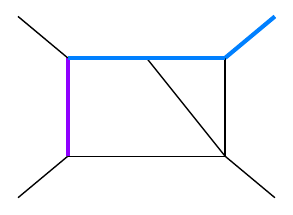}}}
&\xrightarrow[]{6\times\text{cut}}
\int \frac{dz}{
    \sqrt{\alpha(z)\beta(z)}}, \\ 
\vcenter{\hbox{\includegraphics[scale = 1]{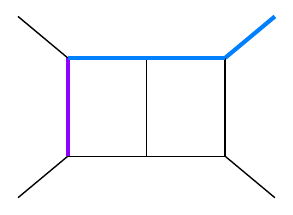}}}
&\xrightarrow[]{7\times\text{cut}}
\int \frac{dz}{
    (m_t^2-s)z\sqrt{\beta(z)}},
\end{aligned}
\label{eq:maximal cut}
\end{equation}
where
\begin{equation}
\begin{aligned}
    \alpha(z) &= z \left(z-4 m_t^2\right),  \\
    \beta(z)  &= s^2 z^2+2 s \left[(s-m_t^2)t+(s+2 t-m_t^2)m_W^2 \right]z+\left(m_t^2-s\right){}^2 \left(m_W^2-t\right){}^2.
\end{aligned}
\end{equation}
One can immediately obtain an elliptic curve defined by the following polynomial equation,
\begin{equation}
    y^2 = \alpha(z)\beta(z),
\label{eq:elliptic curve}
\end{equation}
with the four roots of the quartic polynomial on the right-hand side being
\begin{equation}
    z_1 = 0,\quad z_2 = 4m_t^2,\quad z_{3,4} = \frac{(s-m_t^2)t+(s+2 t-m_t^2)m_W^2\pm 2\sqrt{\lambda} }{s},
\end{equation}
and
\begin{equation}
    \lambda=t m_W^2 \left(s+t-m_t^2\right) \left(s+m_W^2-m_t^2\right).
\end{equation}
It turns out that the first integration evaluates to the elliptic integral and the second to the logarithm. Indeed, we have $\epsilon$ factorized out in the homogeneous part of the differential equation that the top sector satisfies. Its elliptic dependence comes from the sub-sector in which the first integral lives. Then one can expect that in this family, we are not able to construct a UT basis like what we do in families without elliptic sectors.

Another interesting observation from the first maximal cut in \eqref{eq:maximal cut} is that the elliptic curve \eqref{eq:elliptic curve} will degenerate if $m_W^2 \rightarrow 0$,
\begin{equation}
    y^2|_{m_W^2\rightarrow0} = z \left(z-4 m_t^2\right)\left[(m_t^2-s)t+sz\right]^2,
\end{equation}
and the maximal cut becomes a logarithmic function. Taking such a limit, in principle, should be done with care. However, this implies that it might be possible to perform an expansion of $m_W^2$ over $m_t^2$ on the differential equations and then transform them into the canonical form (see Ref.\cite{Gao:2023bll,DiMicco:2019ngk,Wang:2021rxu,Wang:2020nnr,Wang:2019fxh,Xu:2018eos} about the mass-expansion idea). Such an approximation is, to some extent, justified since $m_W^2/m_t^2 \sim 1/4$. Another benefit is that the expansion will rationalize some square roots in the canonical differential equations of some sub-sectors. According to our preliminary calculation, it works, at least in the leading order, even for the families from the third column in Fig. \ref{fig:Feynman diagrams}, which, in our opinion, includes the most complicated diagrams, considering the most number of internal massive propagators. This idea makes it very hopeful to bypass the elliptic issues and to know more about the other diagrams under the expansion.

\section{Summary}
\label{sec:summary}
 In this paper, we studied the two-loop Feynman integrals for the non-factorizable $t$-channle single top quark production, using the methods of UT basis and canonical differential equations. The non-factorizable Feynman diagrams in this process can be reduced towards 9 different integral families, as shown in Fig. \ref{fig:Feynman diagrams}. These diagrams are categorized by their topologies and their numbers of massive internal propagators. According to the computation method we are using, the possibility of constructing a complete UT basis relies on whether this family is free of elliptic sectors or not.
In the 9 diagrams concerned by this paper, we found 3 families with 1 internal massive propagator are indeed free of elliptic sectors. For these families, we provided full analytic expressions of the UT basis in terms of GPL functions. Using these, one can acquire the analytic expressions of the Laporta basis by acting a transformation matrix (see attached files) on the UT results. 
For families that consist of elliptic sectors, it is difficult to build a complete UT basis and full canonical differential equations. However, it is still worth a try to apply methods of UT and canonical differential equations of sub-sectors or expansion on the variables, say $\frac{m_W^2}{m_t^2}$, if they are free of square roots which are not simultaneously rationalizable. Thus, though we cannot express the whole family in form of GPL functions, it is hopefully we can express part of integrals in terms of GPL functions. This helps us to rebuild the information that we are interested in the whole family.

\acknowledgments

We acknowledge Yang Zhang for important aids and discussions. Zihao Wu is supported by the NSF of China through Grant No. 11947301, 12047502, 12075234, and 12247103. Ming-Ming Long is supported by the Deutsche Forschungsgemeinschaft (DFG, German Research Foundation) under grant 396021762 - TTR 257. Zihao Wu also acknowledges Ming-Ming Long for his agreement to re-order the author names (see below).

\textbf{Statement about author ranking}:
The authors of this paper, Ming-Ming Long and Zihao Wu both contribute mainly to this work. Conventionally, the author names should have been ranked in a lexicographic order. The authors have agreed to break this convention in this paper, in order to meet Zihao Wu's PhD graduation requirement of his university, which requires a certain number of first-authored research papers.

\appendix

\section{Attached files}\label{appendix: attached files}
Our paper is attached with some files in the folder ``attachment\_files''. They are some important results of our computation. The files are in Mathematica-readable format, and can be read by the Mathematica command ``Get''. In this section, we show the content of the files. 

In subfolders ``attachment\_files/db1'' and ``attachment\_files/lxb1'', we put in the results of the planar double box diagram and the left-crossed double box diagram, respectively. The contents of each file are shown in Table \ref{tab: attached files - db1 and lxb1}, in both subfolders. 

\begin{table}[h]
    \centering
    
    \begin{tabular}{|c|c|}
    \hline
         file name & content\\
         \hline
         analytic\_UT.txt&The analytic expressions of the UT integrals $\bar{I}_i$\\
         \hline
         Atilde.txt&The $\small{\Tilde{A}}$ matrix\\
         \hline
         
         DEMI.txt&The master integrals \& their differential equation matrices\\
         
         \hline
         DEUT.txt&The  UT basis \& their differential equation matrices\\
         \hline
         letterDef.txt&Expressions of the symbol letters appearing in $\Tilde{A}$ \\
         \hline
         MI.txt&The definition of Laporta master integrals \\
         \hline
         MI2UT.txt&The transform matrix from  Laporta master integrals to UT\\
         \hline
         numericUT\_AMF.m&$(\bar{I}_i)_{\text{AMF}}$ on numerical point \eqref{eq:numeric check point} \\
         \hline
         numericUT\_GPL.m&$(\bar{I}_i)_{\text{GPL}}$ on numerical point \eqref{eq:numeric check point} \\
         \hline
         UT2MI.txt&The transform matrix from UT to Laporta master integrals \\
         \hline

    \end{tabular}
    \caption{The attached files in folder db1 and lxb1.}
    \label{tab: attached files - db1 and lxb1}
\end{table}
In the subfolder ``attachment\_files/rxb1'', we put in the results of the right-crossed double box diagram. The contents of each file are shown in Table \ref{tab: attached files - rxb1}. We did not contain the results of differential equations because they are oversized. You can derive them from the existing files.

\begin{table}[h]
    \centering
    
    \begin{tabular}{|c|c|}
    \hline
         file name & content\\
         \hline
         analytic\_UT.txt&The analytic expressions of the UT integrals $\bar{I}_i$\\
         \hline
         Atilde.txt&The $\small{\Tilde{A}}$ matrix\\
         \hline
         letterDef.txt&Expressions of the symbol letters appearing in $\Tilde{A}$ \\
         \hline
         MI.txt&The definition of Laporta master integrals \\
         \hline
         MI2UT.txt&The transform matrix from  Laporta master integrals to UT\\
         \hline
         numericUT\_AMF.m&$(\bar{I}_i)_{\text{AMF}}$ on numerical point \eqref{eq:numeric check point} \\
         \hline
         numericUT\_GPL.m&$(\bar{I}_i)_{\text{GPL}}$ on numerical point \eqref{eq:numeric check point} \\
         \hline
         UT.txt&The UT basis \\
         \hline
         UT2MI.txt&The transform matrix from UT to Laporta master integrals \\
         \hline

    \end{tabular}
    \caption{The attached files in folder rxb1.}
    \label{tab: attached files - rxb1}
\end{table}

Notice that, in the above files, there are lists (vectors) or matrices with respect to the corresponding UT basis or Laporta master integral basis. The corresponding ordering is the same as the definition of the bases. For example, the file "analytic\_UT.txt" contains the list of results like $\{\bar{I}_1,\bar{I}_2,\cdots\}$, in the same order as that the UT integrals are defined. The "UT2MI.txt" contains the transformation matrix $T$ (after IBP reduction) such that
\begin{equation}
    I_i=T_{ij}J_j,
\end{equation}
where $J_j$ labels the Laporta master basis, with the same order defined in file "MI.txt" (or "DEMI.txt"), and $I_i$ denotes the UT basis, also the same-ordered as other files. Besides, the notations in the result files are slightly different from those in this paper. The differences are listed in Table. \ref{tab: notation differences}.
\begin{table}[h]
    \centering
    
    \begin{tabular}{|c|c|}
    \hline
         notations in the result files & notations in this paper\\
         \hline
         mtt & $m_t^2$\\
         \hline
         mWW & $m_W^2$\\
         \hline
         ep & $\epsilon$\\
         \hline
         W[1],W[2],$\cdots$  & $W_1,W_2,\cdots$\\
         \hline
         x1,y1,z1 & $x_1,y_1,z_1$\\
         \hline

    \end{tabular}
    \caption{Notation differences between attached files and expressions in this paper.}
    \label{tab: notation differences}
\end{table}

\bibliographystyle{JHEP}
\bibliography{bibtex}

\end{document}